\documentclass[11pt]{article}
\usepackage{jheppub} 
\usepackage[normalem]{ulem}
\usepackage[T1]{fontenc}
\usepackage{slashed,amsmath,amsfonts,amssymb,bbm}
\usepackage{mathrsfs,wasysym}
\usepackage{units}
\usepackage{hyperref}
\usepackage{graphicx}
\usepackage{feynmp}
\usepackage{makecell}
\usepackage{booktabs,multirow}
\usepackage[usenames,dvipsnames]{xcolor}

\usepackage{lineno}

\graphicspath{{./Figures/}}

\newcommand{\e}{\varepsilon}
\newcommand{\g}{\gamma}
\newcommand{\s}{\sigma}
\renewcommand{\to}{\rightarrow}
\DeclareMathOperator{\Br}{Br}
\newcommand{\Lag}{\mathscr{L}}
\newcommand{\de}{\partial}

\newcommand{\st}{s_\theta}
\newcommand{\ct}{c_\theta}
\newcommand{\cW}{c_{\widetilde W}}
\newcommand{\cB}{c_{\widetilde B}}
\newcommand{\cG}{c_{\widetilde G}}

\newcommand{\MGMCatNLO}{\textsc{MadGraph5\_aMC@NLO}}
\newcommand{\MG}{\textsc{MadGraph5}}

\subheader{\rm IFT-UAM/CSIC-22-7\\VBSCAN-PUB-01-22}

\title{Nonresonant Searches for Axion-Like Particles in Vector Boson Scattering Processes at the LHC}
\author[a,b]{J.~Bonilla,}
\author[c]{I.~Brivio,}
\author[a,b]{J.~Machado-Rodr\'iguez,}
\author[a]{J.~F. de Troc\'oniz}
\affiliation[a]{Departamento de F\'isica Te\'orica, Universidad Aut\'onoma de Madrid, Cantoblanco,\\ E-28049, Madrid, Spain}
\affiliation[b]{Instituto de F\'isica Te\'orica IFT-UAM/CSIC, 
Cantoblanco, E-28049, Madrid, Spain}
\affiliation[c]{Institut f\"ur Theoretische Physik, Universit\"at Heidelberg, Philosophenweg 16,\\ D-69120 Heidelberg, Germany}
\emailAdd{jesus.bonilla@uam.es}
\emailAdd{brivio@thphys.uni-heidelberg.de}
\emailAdd{jonathan.machado@uam.es}
\emailAdd{jorge.troconiz@uam.es}

\abstract{
We propose a new search for Axion-Like Particles (ALPs), targeting Vector
Boson Scattering (VBS) processes at the LHC. We consider nonresonant ALP-mediated VBS, where the ALP participates as an off-shell mediator. This process occurs whenever the ALP is too light to be produced resonantly, and it takes advantage of the derivative nature of ALP
interactions with the electroweak Standard Model bosons. We study the
production of $ZZ$, $Z\g$, $W^\pm \g$, $W^\pm Z$ and $W^\pm W^\pm$ pairs
with large diboson invariant masses in association with two jets.
Working in a gauge-invariant framework, upper limits on ALP couplings to
electroweak bosons are obtained from a reinterpretation of Run~2
public CMS VBS analyses. The constraints inferred on ALP couplings to $ZZ$, $Z\gamma$ and $W^\pm W^\pm$ pairs are very competitive for ALP masses up to
$\unit[100]{GeV}$. They have the advantage of being independent of the ALP coupling to gluons and of the ALP decay width. Simple projections for LHC Run~3 and HL-LHC are also
calculated, demonstrating the power of future dedicated analyses at ATLAS
and CMS.
}

\begin{document}
\maketitle

\section{Introduction}

Axion-Like Particles (ALPs) constitute a particularly attractive class of hypothetical particles, that are predicted in a variety of Standard Model (SM) extensions, ranging from invisible axion models~\cite{Peccei:1977hh, Peccei:1977ur, Weinberg:1977ma,Wilczek:1977pj,Kim:1979if,Shifman:1979if,Dine:1981rt,Zhitnitsky:1980tq} to string theory~\cite{Cicoli:2013ana}. They are defined as the pseudo-Goldstone bosons of a generic, spontaneously broken global symmetry, that is restored only at energy scales much higher compared to the electroweak (EW) one. Besides the Peccei-Quinn symmetry, typical examples are the lepton number~\cite{Gelmini:1980re,Langacker:1986rj,Ballesteros:2016xej} or flavor symmetries~\cite{Wilczek:1982rv,Ema:2016ops,Calibbi:2016hwq}. Being pseudo-Goldstone bosons, ALPs are pseudo-scalar particles, singlets under the SM gauge groups, and  naturally much lighter than the beyond-SM (BSM) sector they originate from. 
As a consequence, they are most conveniently studied in an Effective Field Theory (EFT) framework, constructed as an expansion in inverse powers of the ALP characteristic scale $f_a$.

At the leading order, the ALP EFT only includes very few parameters (up to flavor indices). Nevertheless, the ranges allowed \emph{a priori} for both the ALP mass $m_a$ and scale $f_a$ are  extremely vast, spanning several orders of magnitude. As a result, the phenomenology of ALPs is one of the richest in particle and astroparticle physics. This peculiarity, together with their ubiquity in BSM models, has recently brought this class of particles into the spotlight, stimulating enormous theoretical and experimental advancements. A plethora of experiments searching for ALPs in different regimes and exploiting very diversified techniques are either already taking  data or scheduled to do so in the next decade, see e.g.~\cite{Beacham:2019nyx,Agrawal:2021dbo} for recent reports. 

Most of these experiments are sensitive to ALPs coupling to photons, electrons or gluons. 
ALP interactions with the massive gauge bosons, on the other hand, are harder  to access:  at present, they can only be probed indirectly via loop corrections to low-energy processes~\cite{Izaguirre:2016dfi,Bauer:2017ris,Alonso-Alvarez:2018irt,Gavela:2019wzg,Ebadi:2019gij,Ertas:2020xcc,Gori:2020xvq,Kelly:2020dda,Galda:2021hbr,Bonilla:2021ufe,Bauer:2021mvw} or directly at colliders. 
At the LHC, depending on the ALP mass and decay width, ALP-gauge interactions can be probed in $V+$ALP associated production processes, with $V=\g,Z,W^\pm$ and the ALP either escaping detection~\cite{Mimasu:2014nea,Brivio:2017ije} or decaying resonantly~\cite{Craig:2018kne,Wang:2021uyb,Ren:2021prq}, in resonant ALP decays into diboson pairs~\cite{Craig:2018kne}, or in nonresonant processes where the ALP enters as an off-shell mediator.
The latter were first studied in the context of inclusive
diboson production at the LHC, where the ALP appears in $s$-channel, being produced via gluon fusion~\cite{Gavela:2019cmq}. These channels are sensitive to the product of the ALP coupling to gluons with the relevant coupling to dibosons and probe previously unexplored areas of the ALP parameter space. Moreover, the nonresonant cross sections and kinematical distributions are found to be independent of the ALP mass from arbitrarily light masses up to masses of the order of $\unit[100]{GeV}$~\cite{Gavela:2019cmq}. The experimental strategy is to look for deviations with respect to SM expectations in the tails of the bosons transverse momenta or diboson mass distributions. ALP coupling limits derived from reinterpretations of CMS and ATLAS Run~2 measurements were presented in Refs.~\cite{Gavela:2019cmq,Carra:2021ycg}, while the CMS Collaboration has recently published a dedicated search for nonresonant ALP-mediated $ZZ$ production in semileptonic final states at the LHC~\cite{CMS:2021xor}. 

\begin{figure}[t]\centering
\includegraphics[width=5.cm, page=2]{VBS_ALP_diagrams.pdf}
\includegraphics[width=5.cm, page=1]{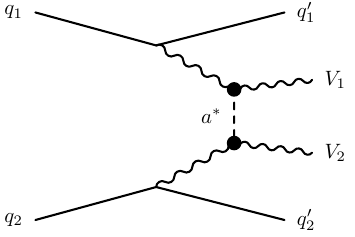}
\includegraphics[width=5.cm, page=3]{VBS_ALP_diagrams.pdf}
\caption{Feynman diagrams for ALP contributions to a generic process $q_1 q_2 \to q'_1 q'_2 V_1 V_2$. Fermion lines represent both quarks and antiquarks. In the last diagram, the final state quarks can be emitted from any of the outgoing bosons.}\label{fig.ALP_diagrams}
\end{figure}

In this paper we study for the first time nonresonant ALP signals in EW Vector Boson Scattering (VBS) processes at the LHC (see~\cite{Covarelli:2021gyz} for a review). We focus on channels containing massive EW bosons: ALP EW VBS processes with the ALP going to a photon pair were studied in Ref.~\cite{Florez:2021zoo} for the LHC, in Ref.~\cite{Yue:2021iiu} for CLIC and in Ref.~\cite{Liu:2021lan} for the EIC.  Fig.~\ref{fig.ALP_diagrams} depicts the leading order Feynman diagrams for ALP-mediated EW production of $q_1 q_2 \to q'_1 q'_2 V_1 V_2$. The two jets in the final state, $q'_1$ and $q'_2$, are required to have a large invariant mass and to be well separated in rapidity. 
These processes are particularly convenient for a number of reasons: first, they allow us to access the couplings of the ALP to EW bosons independently of the coupling to gluons. At the same time, the richness of VBS in terms of different final states helps  constraining the parameter space from multiple complementary directions.

Searching for signals beyond the SM in VBS final states is a major goal for the ATLAS and CMS experiments and both collaborations have recently reported Run~2 measurements of these processes~\cite{CMS:2017fhs,ATLAS:2018mxa,CMS:2019uys,Sirunyan:2019der,ATLAS:2019thr,ATLAS:2019cbr,ATLAS:2019qhm,CMS:2020ioi,ATLAS:2020nlt,Sirunyan:2020gyx,Sirunyan:2020alo,Sirunyan:2020azs,Sirunyan:2020gvn,CMS:2021gme,CMS:2021mjn,CMS:2021rmh}. These analyses allow us to perform a first comparison of the ALP VBS predictions to the data, a calibration of the available simulation tools and a calculation of educated predictions for higher LHC luminosities. 
Moreover, nonresonant searches are generally expected to become more and more competitive during the upcoming LHC runs. They will benefit, on the one hand, from the large increase in the accumulated statistics and, on the other, from the technological developments currently driven by studies of the Standard Model EFT (SMEFT) formalism, that are encouraging a global, comprehensive approach to new physics searches.
Interestingly, while SMEFT analyses rely on the assumption of new particles being too heavy to be produced on-shell, nonresonant ALP searches target the opposite limit, i.e. where the ALP is \emph{too light} to decay resonantly. In this way, they provide access to parameter space regions complementary to those probed in other LHC searches. In particular, compared to resonant or large-missing-momentum processes, they require only very minimal assumptions on the ALP decay width.

The manuscript is organized as follows: the theoretical framework adopted is defined in Section~\ref{sec:Lagrangian}. In Sec.~\ref{sec:VBS} we discuss the general characteristics of nonresonant ALP EW VBS production. The details of the ALP VBS simulation and analysis are explained in Sec.~\ref{sec:numerics}. We first extract current constraints on ALP-gauge interactions from measurements of differential VBS observables published by the CMS Collaboration, and subsequently estimate projected limits for the LHC Run~3 and for the High Luminosity (HL-LHC) phase. The results are presented in Sec.~\ref{sec:results}. In Section~\ref{sec:comparison} we compare them to other existing constraints. In Sec.~\ref{sec:conclusions} we conclude.

\section{The ALP Effective Lagrangian}\label{sec:Lagrangian}

We define the ALP  $a$ as a pseudo-scalar state whose interactions are either manifestly invariant under shifts $a(x)\to a(x) + c$ (as befits its Goldstone origin) or generated via the chiral anomaly. 
Adopting an EFT approach, all ALP interactions are weighted down by inverse powers of the characteristic scale $f_a\gg m_a$, that is unknown and naturally close to the mass scale of the heavy sector the ALP originates from. We implicitly assume $f_a\gg v\simeq \unit[246]{GeV}$ and require all ALP interactions to be invariant under the full SM gauge group. We neglect CP violating terms and ALP-fermion interactions. The latter only give highly suppressed contributions at tree-level, as their physical impact is always proportional to the mass of the fermion itself and only light fermions appear in LO VBS  diagrams.

The SM is then extended by the Lagrangian~\cite{Georgi:1986df, Choi:1986zw}
\begin{equation}
\label{eq.L_ALP}
\Lag_{ALP} = \frac{1}{4}\de_\mu a\de^\mu a - \frac{m_a^2}{2}a^2 
- \cB\frac{a}{f_a} B_{\mu\nu}   \tilde B^{\mu\nu}
- \cW\frac{a}{f_a} W^i_{\mu\nu} \tilde W^{i\mu\nu}
- \cG\frac{a}{f_a} G^A_{\mu\nu} \tilde G^{A\mu\nu}\,,
\end{equation}
that contains a complete and non-redundant set of dimension-5 bosonic operators.\footnote{One more bosonic dimension~5 operator could be written down, namely $O_{a\Phi} = \de^\mu a(\Phi^\dag i\overleftrightarrow{D_\mu} \Phi)$, where $\Phi$ the Higgs doublet. However, this operator can be fully traded for ALP-fermion terms via the Higgs equations of motion~\cite{Georgi:1986df}. Therefore, its impact on VBS processes is negligible.}
Here $B_\mu,W^i_\mu,G^A_\mu$ denote the bosons associated to the $U(1),\, SU(2)$ and $SU(3)$ gauge symmetries of the SM, respectively. The associated coupling constants will be denoted by $g',g, g_s$. Unless otherwise specified, we will use $i,j,k$ and $A,B,C$ to denote isospin and color indices. 
Covariant derivatives are defined with a minus sign convention, such that $W_{\mu\nu}^i = \de_\mu W^i_\nu-\de_\nu W^i_\mu + g \epsilon^{ijk} W^j_\mu W^k_\nu$ and analogously for gluons.
Dual field strengths are defined as $\tilde X_{\mu\nu} = \frac{1}{2} \e_{\mu\nu\rho\sigma} X^{\rho\sigma}$.

In the analysis presented below, we only consider EW ALP contributions to the VBS processes, while we neglect those containing ALP-gluon interactions, which is tantamount to setting $\cG=0$.
This is a very good approximation for the $W^\pm \g$, $W^\pm Z$ and same-sign $W^\pm W^\pm$ channels where the ALP QCD contribution is absent at tree level. 
For the $ZZ$ and $Z \g$ channels, an ALP QCD contribution is present in principle. However, the ALP QCD component is reduced by requiring consistency with the limits obtained in~\cite{Gavela:2019cmq,Carra:2021ycg,CMS:2021xor}, the rejection of the VBS selection cuts and the large diboson invariant masses involved.
In particular, for values of the EW couplings $\gtrsim \unit[1]{TeV}^{-1}$, the theoretical prediction is dominated by the pure EW ALP signal, with a smaller contribution from the pure QCD ALP signal.
Here, both the EW and QCD ALP signal components are positive and their interference is subdominant. This rules out the possibility of cancellations between the ALP EW and QCD components, and implies that the final bounds for $\cW/f_a$ and $\cB/f_a$ for $\cG=0$ are conservative.

It is then safe to restrict the parameter space to the four ALP couplings to the electroweak gauge bosons. In unitary gauge, they are usually parameterized as
\begin{equation}\label{eq.Lalp_physical}
\Lag_{ALP, EW} = 
-\frac{g_{a\g\g}}{4}  aF^{\mu\nu}\tilde F_{\mu\nu}
-\frac{g_{a\g Z}}{4}  aZ^{\mu\nu}\tilde F_{\mu\nu}
-\frac{g_{aZZ}}{4}  aZ^{\mu\nu}\tilde Z_{\mu\nu}
-\frac{g_{aWW}}{2}  aW^{+\mu\nu}\tilde W^-_{\mu\nu}\,,
\end{equation}
with $F_{\mu\nu}, Z_{\mu\nu}, W^\pm_{\mu\nu}$ are the field strengths of the photon, $Z$ and $W^\pm$ bosons respectively,
\begin{align}
g_{a\g\g} &=\frac{4}{f_a}(\st^2\cW+\ct^2\cB)\,,
&
g_{a\g Z} &=\frac{4}{f_a}s_{2\theta}(\cW-\cB)\,,
\\
g_{aZZ} &=\frac{4}{f_a}(\ct^2\cW+\st^2\cB)\,,
&
g_{aWW} &=\frac{4}{f_a}\cW\,,
\end{align}
and $\st,\ct$ the sine and cosine of the weak mixing angle.
For later convenience, we also define
\begin{equation}
\label{g_agg}
    g_{agg} = \frac{4}{f_a} \cG\,.
\end{equation}

\section{General Characteristics of ALP-mediated EW VBS Production}\label{sec:VBS}

We consider the production of $ZZ$, $Z\g$, $W^\pm \g$, $W^\pm Z$ and same-sign $W^\pm W^\pm$ pairs in association with two forward jets. These five VBS channels are those for which differential measurements of the diboson invariant mass (or transverse mass) have been reported by the CMS Collaboration, using data collected at the LHC Run~2. At parton level we treat them, for simplicity, as $2\to 4$ scatterings, with either photons or weak bosons in the final state. As described in Sec.~\ref{sec:numerics}, the weak bosons are decayed to leptons at a later stage.

ALPs give EW contributions to these processes via the diagram topologies shown in Fig.~\ref{fig.ALP_diagrams}. All of them necessarily present two insertions of ALP operators, leading to amplitudes that scale as $f_a^{-2}$, and cross sections of order $f_a^{-4}$.
A generic VBS cross section, including both SM and EW ALP contributions, has the structure
\begin{equation}
\label{eq.polynomial}
\begin{aligned}
\s_{ALP} &= \s_{SM} 
+ \frac{1}{f_a^2}\, \s_{\rm interf.}
+ \frac{1}{f_a^4}\, \s_{\rm signal}\,,
\\
\s_{\rm interf.} &=\cB^2\,\s_{B2} + \cW^2 \,\s_{W2} + \cB \cW \,\s_{BW} \,,
\\[2mm]
\s_{\rm signal} &= \cB^4 \,\s_{B4} + \cW^4 \,\s_{W4} + 
\cB^2 \cW^2 \,\s_{B2W2}+
\cB^3 \cW \,\s_{B3W}+
\cB \cW^3 \,\s_{BW3}\,,
\end{aligned}
\end{equation}
where all the $\sigma_i$ quantities can be evaluated numerically from the simulations.
This structure holds after selection cuts.
Not all processes receive contributions from all terms in this polynomial expansion: the dependence is summarized in Tab.~\ref{tab.channels}. The pattern observed can be easily explained: all processes with a $W$ boson in final state require an insertion of $g_{aWW}\sim\cW / f_a$. Pure $\cB$ contributions are then absent, which means that these channels cannot constrain the ALP parameter space along the $\cB$ axis. Same-sign $W^\pm W^\pm$ production represents an extreme case where $\cB$ does not enter at all.
Explicit expressions of $\s_{\rm interf.}$, $\s_{\rm signal}$ are given in App.~\ref{app:polynomial} for the integrated cross-section of each channel, calculated after the selection cuts, see Sec.~\ref{sec:numerics}.

\begin{table}[t]
\centering\renewcommand{\arraystretch}{1.3}
\begin{tabular}{>{$}l<{$}|*3c|*5c}
\toprule
\text{Process}& $\cB^2$& $\cW^2$ & $\cB\cW$ & $\cB^4$ & $\cW^4$ & $\cB^2\cW^2$ & $\cB^3\cW$ & $\cB\cW^3$
\\\midrule
pp\to jj Z Z & 
\checkmark& \checkmark& \checkmark& \checkmark& \checkmark& \checkmark& \checkmark& \checkmark\\
pp\to jj Z \g& 
\checkmark& \checkmark& \checkmark& \checkmark& \checkmark& \checkmark& \checkmark& \checkmark\\
pp\to jj W^\pm \g&
& \checkmark& \checkmark& & \checkmark& \checkmark& & \checkmark\\
pp\to jj W^\pm Z &
& \checkmark& \checkmark& & \checkmark& \checkmark& & \checkmark\\
pp\to jj W^\pm W^\pm & 
& \checkmark& & & \checkmark& & & \\
\bottomrule
\end{tabular}
\caption{List of VBS processes considered in this work. For each, we indicate which terms in the polynomial dependence on $\cW$, $\cB$ (Eq.~\eqref{eq.polynomial}) are present in the parameterization of the ALP signal.}
\label{tab.channels}
\end{table}

Among the diagrams shown in Fig.~\ref{fig.ALP_diagrams}, the first, where the ALP is exchanged in $s$-channel, only contributes to VBS with $ZZ$ and $Z\gamma$ final states.\footnote{The $s$-channel diagram contributes also to VBS with opposite-sign WW or diphoton final states. However, these channels are not considered here.} The second, with the ALP in $t$-channel, is relevant for all VBS processes, and it is the only one contributing to $W^\pm\g,W^\pm Z$ and $W^\pm W^\pm$. Finally, the third topology is triboson-like: although these diagrams were included for consistency in our calculation,  we have verified that their contribution is efficiently suppressed with a cut on the dijet invariant mass $M_{q'_1 q'_2} > \unit[120]{GeV}$.

\begin{figure}[t]
\centering
\includegraphics[width=7.4cm]{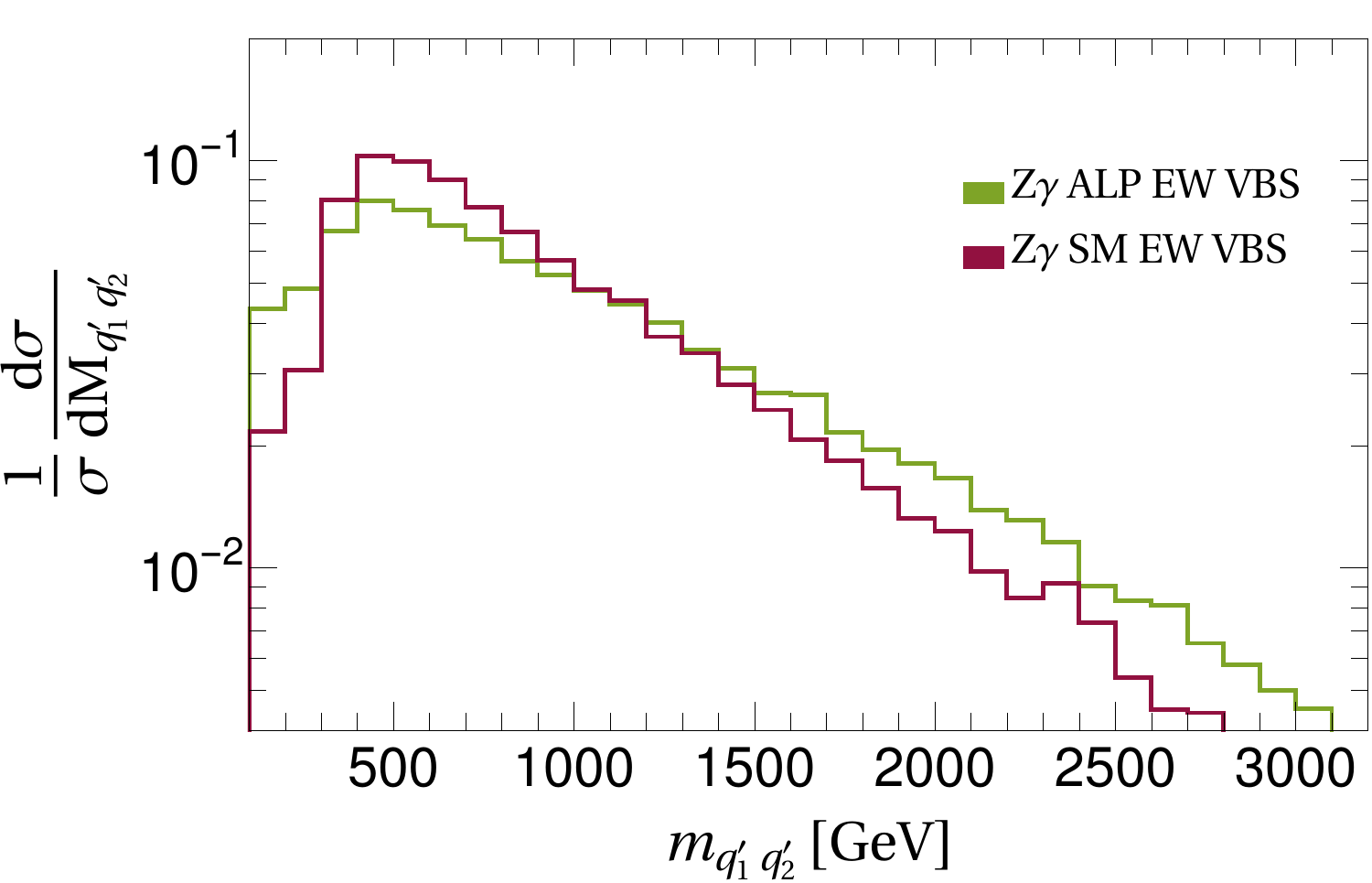}
\hfill
\includegraphics[width=7.4cm]{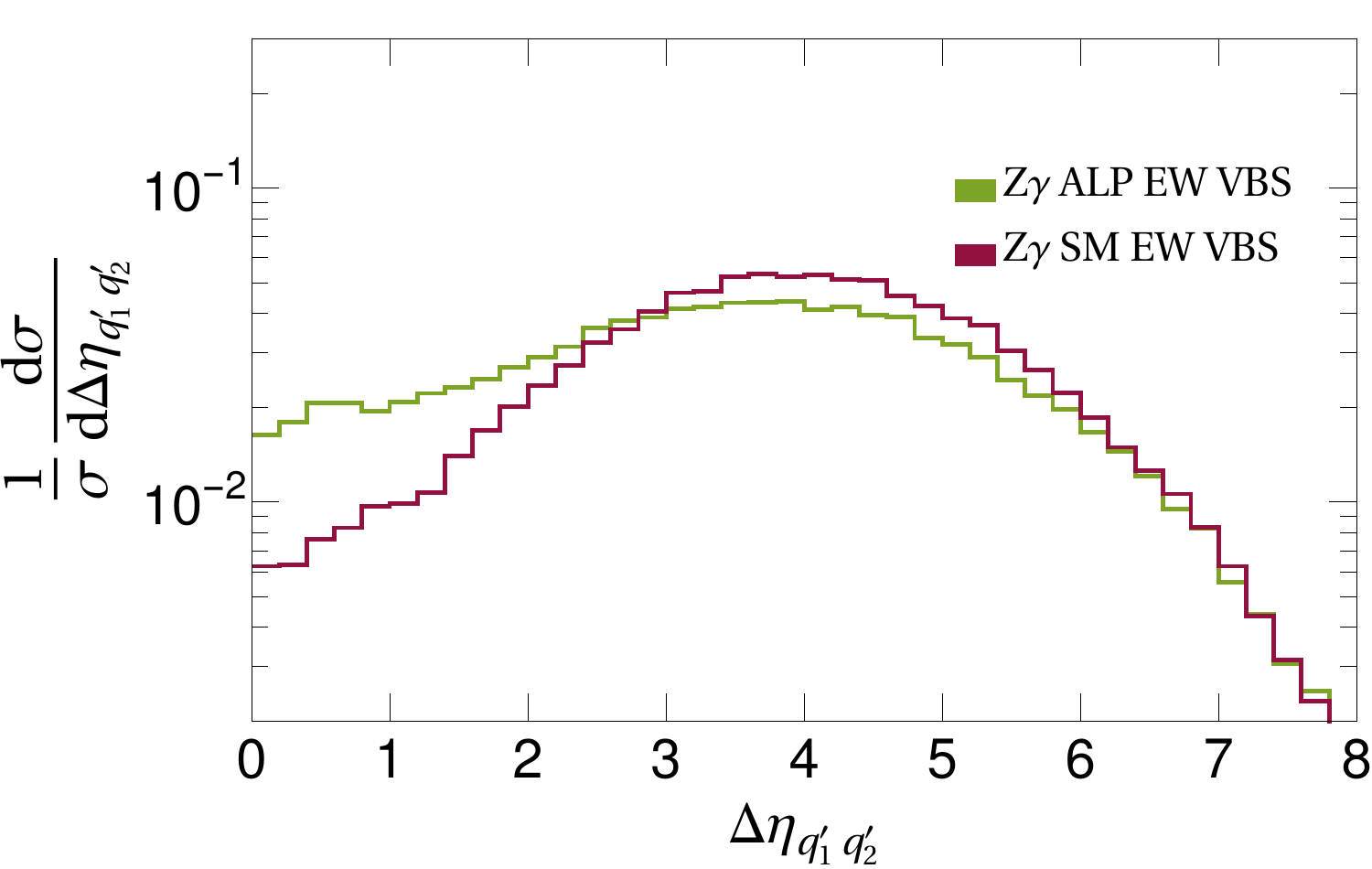}\\
\includegraphics[width=7.4cm]{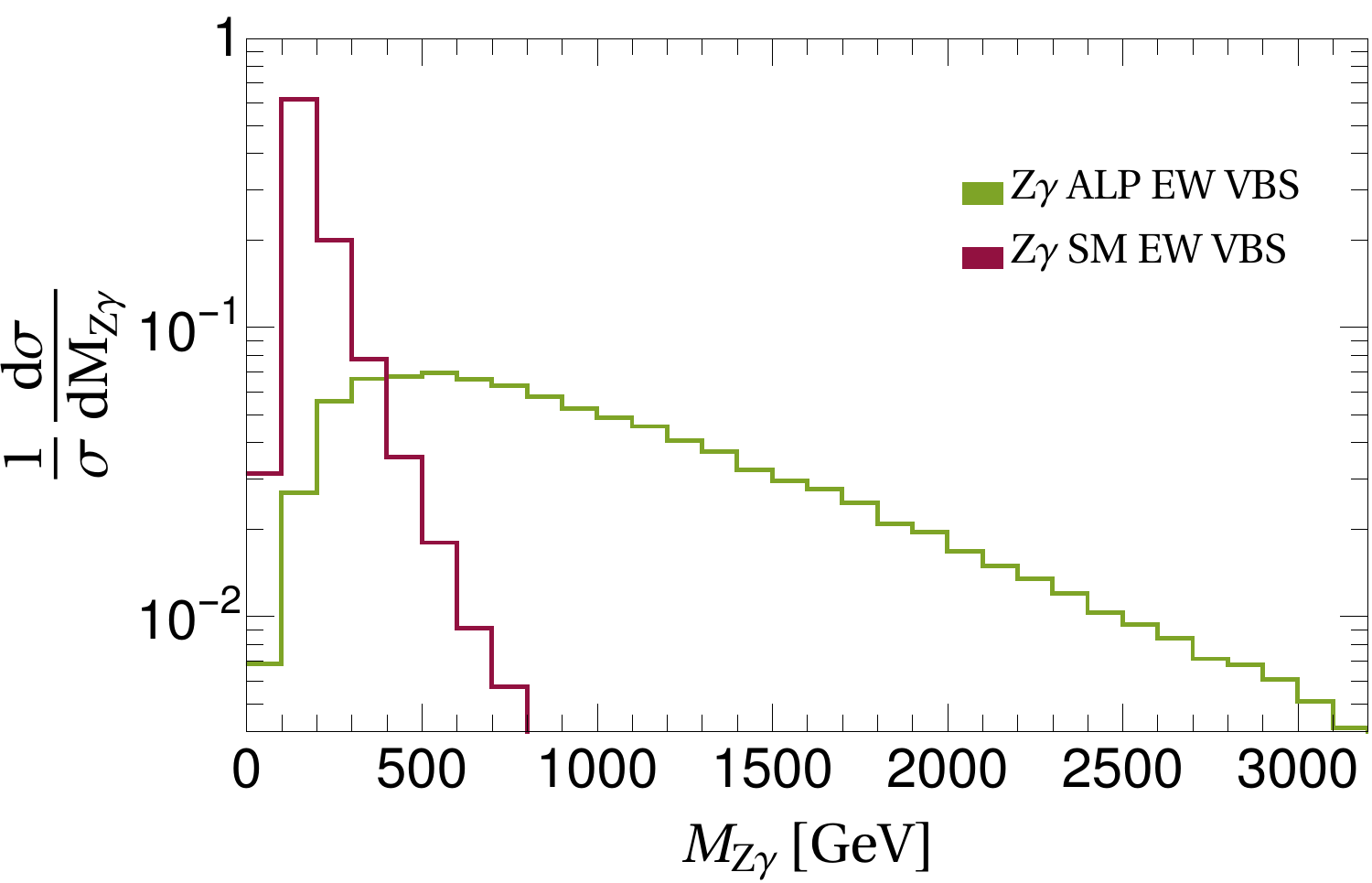}
\caption{Normalized parton level distributions of the dijet invariant mass $M_{q'_1 q'_2}$, jet pseudo-rapidity separation $\Delta \eta_{q'_1 q'_2}$ and diboson invariant mass $M_{Z\g}$ for ALP EW VBS (green) and SM EW VBS (red) $Z\g$ production. The ALP curves include the pure signal and ALP-SM interference contributions, computed  for $m_a=\unit[1]{MeV}$ and $\cW/f_a=\cB/f_a=\unit[1]{TeV^{-1}}$.}
\label{fig:variables}
\end{figure}

We take the ALP to be too light for any of the $V_1 V_2$ pairs to be produced resonantly. As a consequence, the ALP is always off-shell and its propagator acts as a suppression of the scattering amplitudes. However, this effect is overcompensated by the momentum enhancement induced by the ALP interaction vertices. The net result is that the ALP-mediated cross section falls more slowly with the diboson invariant mass of the boson pair $M_{V_1 V_2}$ than the SM backgrounds. Fig.~\ref{fig:variables} shows a parton-level comparison of the dijet invariant mass $M_{q'_1 q'_2}$, jet pseudo-rapidity separation $\Delta \eta_{q'_1 q'_2}$ and diboson invariant mass $M_{Z\g}$ distributions for ALP EW VBS and SM EW VBS $Z\g$ production. The ALP curves include the pure signal and ALP-SM interference contributions, computed  for $m_a=\unit[1]{MeV}$ and $\cW/f_a=\cB/f_a=\unit[1]{TeV^{-1}}$.
The dijet distributions are qualitatively similar, and dijet selection criteria designed to measure the SM EW VBS component should work efficiently for the ALP case as well. On the other hand, the very different tails of the diboson invariant mass distributions allow discrimination of the two processes for $M_{Z\g} \gtrsim \unit[500]{GeV}$.
This general behavior holds for all ALP EW VBS final states, independently of the presence or absence of $s$-channel ALP Feynman diagrams.

\subsection{Comments on the EFT power counting} Before discussing the details of the numerical analysis, a few comments on the validity of the EFT approach are in order. 
In particular, concerns might be raised about the fact that an ALP signal of $\mathcal{O}(f_a^{-4})$ is extracted from a Lagrangian defined at $\mathcal{O}(f_a^{-1})$.
First of all, it should be noted that, because two insertions of ALP operators are always required in order to generate corrections to SM processes, the dimension-5 Lagrangian does provide complete VBS predictions up to $\mathcal{O}(f_a^{-2})$. However, it is indeed possible for $d\geq 6$ ALP operators to induce further contributions at $\mathcal{O}(f_a^{-3},f_a^{-4})$ that are neglected in this work.
Specifically, at tree-level, these missing terms can be exclusively corrections to $\s_{\rm interf.}$ from ALP diagrams containing $d=6$ or $d=7$ operator insertions, while the expression for $\s_{\rm signal}$ is already complete to $\mathcal{O}(f_a^{-4})$. 
Note, in addition, that the parameterization in Eq.~\eqref{eq.polynomial} is complete to quartic order in the parameter space $(\cB/f_a,\cW/f_a)$, i.e. it accounts for \emph{all} contributions  up to $\mathcal{O}(f_a^{-4})$ generated by the $d=5$ Lagrangian\footnote{
This is at variance e.g. with the SMEFT case, where
the square of the $d=6$ amplitude does not contain all $\mathcal{O}(c_6^2/\Lambda^4)$ contributions, because certain $d=6$ operators can induce extra higher-order corrections, e.g. via redefinitions of SM fields or parameters, or via double insertions in a given diagram.}.
These considerations, together with the fact that the analysis presented in the next sections is numerically dominated by $\s_{\rm signal}$ (see e.g. Tab.~\ref{tab:poly}), suggest that the final results of this work would not change significantly if the missing $\mathcal{O}(f_a^{-4})$ terms were restored.\footnote{The results could change significantly only if $d\geq 6$ ALP operators introduced very large kinematic enhancements to $\s_{\rm interf.}$, sufficient to make it competitive with $\s_{\rm signal}$ within the region of sensitivity. Very preliminary considerations about the possible structure of such operators suggest that this is unlikely. }

For these reasons, we deem the Lagrangian in Eq.~\eqref{eq.L_ALP} adequate for the scope of this work and we believe that the resulting constraints on $(\cB/f_a,\cW/f_a)$ are quite solid. We stress that these conclusions are based on considerations made {\it a posteriori}, having evaluated the sensitivity of LHC and HL-LHC VBS searches to $d=5$ ALP couplings. They do not necessarily apply to processes different from VBS or in scenarios with very different sensitivity. 
A systematic and more quantitative assessment of the impact of higher-order ALP operators is left for future work.
Note that this would require, among other things, the definition of a complete and non-redundant ALP operator basis beyond dimension-5, which has not been constructed to date.

\section{ALP-mediated EW VBS Simulation and Analysis}\label{sec:numerics}

In order to understand the potential of the LHC experiments, we perform a reinterpretation of the 
analyses recently published by the CMS Collaboration studying the production of $ZZ$~\cite{Sirunyan:2020alo}, $Z \gamma$~\cite{CMS:2021gme}, $W^{\pm}\gamma$~\cite{Sirunyan:2020azs}, $W^{\pm}Z$~\cite{Sirunyan:2020gyx} and same-sign $W^{\pm}W^{\pm}$~\cite{Sirunyan:2020gyx} bosons in association with two jets. All channels use leptonic (electron and muon) decays of the $W$ and $Z$ bosons in the final state.

The nonresonant ALP-mediated EW VBS diboson signal is simulated with the software \MGMCatNLO\ 2.8.2~\cite{Alwall:2014hca}. Employing the {\tt ALP\_linear} UFO model from~\cite{ALPfeynrules,Brivio:2017ije}, we generate $q_1 q_2 \to V_1 V_2 q'_1 q'_2$ events
at leading order in the ALP and EW couplings and at zeroth order in the QCD coupling, using a 4-flavor-scheme. The parton distribution functions (PDFs) of the colliding protons are given by the NNPDF 3.0 PDF set~\cite{NNPDF:2014otw} for all simulated samples.
Kinematical cuts requiring
\begin{equation}
\begin{aligned}
p_T (q'_{1, 2}) &>\unit[20]{GeV}\,,
&
\eta (q'_{1, 2}) &<6\,,
&
\Delta R (q'_1 q'_2) &>0.1\,,
&
M_{q'_1 q'_2}&>\unit[120]{GeV}\,,
\\
p_T (\g) &>\unit[10]{GeV}\,,
&
\eta (\g) &< 2.5\,,
&
\Delta R (\g q'_{1,2})&> 0.4\,,
& &
\end{aligned}
\label{eq.gen_cuts}
\end{equation}
are imposed at generation level for all VBS processes, except for the $ZZ$ channel where the $M_{q'_1 q'_2}$ cut is removed. The angular separation is defined as $\Delta R = \sqrt{\Delta \eta ^2 + \Delta \phi^2}$, with $\eta$ the parton's pseudorapidity and $\phi$ its azimuthal angle. 
The ALP EW VBS signals are generated fixing $m_a=\unit[1]{MeV}$, $f_a=\unit[1]{TeV}$ and $\Gamma_a=0$. The specific values of the ALP mass and decay width do not have significant consequences in the nonresonant regime, see Sec.~\ref{sec:mass_width}.
We generate separated samples for pure ALP-mediated production and the interference between the ALP and the SM EW VBS production.
As discussed in Sec.~\ref{sec:VBS}, the ALP EW VBS cross sections have a polynomial dependence on the parameters $\cW$ and $\cB$, whose coefficients need to be determined individually. This requires to evaluate the ALP-SM interference at a minimum of three linearly independent points in the $(\cW,\cB)$ plane, and the pure ALP signal at a minimum of five points. This is achieved by exploiting the interaction orders syntax in \MGMCatNLO, used both in independent event generations and with the \MG~reweighting tool~\cite{Artoisenet:2010cn}. For cross-checking purposes, we  consider a redundant set of points, namely
\begin{equation}
\label{eq.points}
\begin{aligned}
p_0&=(1,1), 
&p_1&=(0,2), 
&p_2&=(1,0), \\
p_3&=(1,-1), 
&p_4&=(1,-0.305), 
&p_5&=(1,-3.279)\,,
\end{aligned}
\end{equation}
where $p_0$ lies on the $\cB=\cW$ line, where $g_{a\g Z}=0$; $p_4$ is on the photophobic $\cB=-t_\theta^2\cW$ line, where $g_{a\g\g}=0$; and $p_5$ is on the $\cB=-\cW/t_\theta^2$ line, where $g_{aZZ}=0$. Here $t_\theta$ is the tangent of the Weinberg angle.
We use five of these points to determine the polynomials and verify that the results extrapolated to the sixth point match those from direct simulation. This operation has been repeated on all possible subsets to verify the robustness of the predictions. The resulting polynomial expressions for the total cross sections, obtained after the full simulation and analysis procedure, are reported in App.~\ref{app:polynomial}. These can be employed to estimate the overall normalizations of the ALP signal for all distributions used in the final fits to the data. 
\begin{table}[t]\centering
\renewcommand{\arraystretch}{1.3}
\begin{tabular}{>{$}l<{$}|>{$}r<{$}@{$\,\pm\,$}>{$}l<{$}|>{$}c<{$}|*2{>{$}r<{$}@{$\,\pm\,$}>{$}l<{$}}}
\toprule
\text{Process}& 
\multicolumn{2}{c|}{$\sigma_{SM}$ [fb]}&
\text{Point}& 
\multicolumn{2}{c}{$\sigma_{\rm interf.}$ [fb]} & 
\multicolumn{2}{c}{$\sigma_{\rm signal}$ [fb]}
\\\midrule
pp\to jjZZ& 
98& 1 &
p_0&
-13.5& 0.1 &
42.4& 0.2
\\
&\multicolumn{2}{c}{}& p_4& 
-9.3& 0.1&
18.5& 0.1
\\\midrule
pp\to jjZ\g&
393& 1&
p_0&
0.3& 0.1&
11.1& 0.1
\\
&\multicolumn{2}{c|}{}&
p_4&
-9.1& 0.1& 
20.9& 0.1
\\\midrule
pp\to jjW^\pm \g&
994& 3&
p_0&
4.3& 0.1&
28.7& 0.1
\\
&\multicolumn{2}{c|}{}&
p_4&
1.7& 0.1&
5.4& 0.1
\\\midrule
pp\to jj W^\pm Z& 
386& 1&
p_0&
1.7& 0.1&
18.4& 0.1
\\
&\multicolumn{2}{c|}{}&
p_4&
0.1& 0.1&
23.9& 0.1
\\\midrule
pp\to jj W^\pm W^\pm&
256& 1&
p_0,\,p_4&
-4.0& 0.1&
16.0& 0.1\\
\bottomrule
\end{tabular}
\caption{EW VBS SM background and ALP signal partonic cross sections for $\sqrt{s} = \unit[13]{TeV}$, before decaying the vector bosons and applying only the selection cuts in Eq.~\eqref{eq.gen_cuts}.
The ALP signal cross sections are presented for two benchmark points $p_0$ and $p_4$ defined in Eq.~\eqref{eq.points}. For same-sign $W^\pm W^\pm$, both points give the same results. The reported errors are the statistical errors of the \MGMCatNLO~calculation.}
\label{tab:cross-sections}
\end{table}
The production cross sections at $\sqrt{s} = \unit[13]{TeV}$ for benchmark points $p_0$ and $p_4$ are summarized in Tab.~\ref{tab:cross-sections}.
They have additionally a 11\% systematic uncertainty related to the renormalization and factorization scales and a 4\% systematic uncertainty related to the PDFs.

SM EW VBS diboson background events are generated with \MG~at leading order in the EW couplings and zeroth order in the QCD coupling. This is an irreducible source of background for the analysis. Cross sections at $\sqrt{s} = \unit[13]{TeV}$ are presented in Tab.~\ref{tab:cross-sections}.

For all the simulated samples in the analysis, parton showering, hadronization and decays are described by interfacing the event generators with PYTHIA~8~\cite{SJOSTRAND2015159}. Massive EW bosons $V_1$ and $V_2$ are forced to decay leptonically (electrons and muons). No additional pileup $pp$ interactions were added. All samples were processed through a simulation of the CMS detector and reconstruction of the experimental objects using DELPHES~3~\cite{DELPHES}, including FastJet~\cite{Cacciari:2011ma} for the clustering of anti-$k_T$ jets with a distance parameter of 0.4 (AK4 jets).
The CMS DELPHES card was modified to improve the lepton isolation requirements and to reduce the lepton detection transverse momentum threshold to $\unit[5]{GeV}$.

For the detector-level analysis, we apply the set of requirements designed to constrain anomalous quartic gauge couplings in the CMS publications. The most important cuts are those
imposed on the dijet system, and on the photon transverse momentum if relevant, indicated in Tab.~\ref{tab:analyses}.
Differences between our generation and simulation procedure and the ones used by the CMS experiment are taken into account by comparing the predicted numbers of events after selection cuts for the SM EW VBS processes. In this context, the expected sources of discrepancy are calibration, efficiency or resolution effects in the reconstruction of the experimental observables. 
We observe that all these affect primarily the normalization
and therefore we define a scale factor $\rho$ as the ratio of the number of expected events delivered by our generation and simulation procedure and the number of CMS expected events. We have verified that, after applying this rescaling, our simulation reproduces correctly the relevant kinematic distributions by CMS within the uncertainties.
The same scale factors are then applied to the predictions for pure ALP-mediated EW VBS and ALP-SM interference simulated samples. For each channel, we assign an uncertainty to $\rho$, that stems from the uncertainty on the expected event yield for SM EW VBS production, reported in the CMS publications. 
A systematic uncertainty of $16\%$ on the simulated ALP event yields is assigned, fully correlated across all channels. This is estimated as the average relative error on the scale factors $\rho$.
A summary of the CMS VBS analyses is presented in Tab.~\ref{tab:analyses}: the diboson mass observable, the integrated luminosity, the selection criteria and the normalization scale factor $\rho$.

\begin{table}[t]
\renewcommand{\arraystretch}{1.3}
\begin{tabular}{ccc|l|c}
\toprule
Channel & 
Obs.&
Lum. $[\unit{fb^{-1}}]$ & 
Selection Criteria &
$\rho$ 
\\\midrule
$ZZ$ & $M_{ZZ}$& 137 & 
$M_{jj}>\unit[100]{GeV}$
 &$0.8 \pm 0.1$
\\
$Z\gamma$ & $M_{Z\g}$& 137 &   
$M_{jj}>\unit[500]{GeV},\, \Delta\eta_{jj}>2.5,\, p_T^\g>\unit[120]{GeV}$
& $1.4 \pm 0.2$
\\
$W^\pm\gamma$ & $M_{W\g}$& 35.9 & $M_{jj}>\unit[800]{GeV},\, \Delta\eta_{jj}>2.5,\, p_T^\g>\unit[100]{GeV}$
& $3.1 \pm 0.5$
\\
$W^\pm Z$ & $M_{WZ}^T$& 137  & 
$M_{jj}>\unit[500]{GeV},\, \Delta\eta_{jj}>2.5$
& $1.5 \pm 0.4$
\\
$W^\pm W^\pm$ & $M_{WW}^T$& 137 &
$M_{jj}>\unit[500]{GeV},\, \Delta\eta_{jj}>2.5$
& $1.3 \pm 0.2$
\\
\bottomrule
\end{tabular}
\caption{Summary of the CMS VBS analyses: the diboson mass observable, the integrated luminosity, the most important selection criteria and the normalization scale factor $\rho$.}
\label{tab:analyses}
\end{table}

As anticipated in Sec.~\ref{sec:VBS}, the discrimination between signal and background is based on the diboson mass distributions shown in App.~\ref{app:polynomial}. These include the fully reconstructed diboson invariant masses $M_{ZZ}$ and $M_{Z\g}$; the diboson invariant mass $M_{W^\pm \g}$, where the longitudinal momentum of the neutrino is constrained by the condition $M_{\ell\nu} = m_W$~\cite{Sirunyan:2020azs}; and the diboson transverse masses $M^T_{W^\pm W^\pm}$ and $M^T_{W^\pm Z}$, defined as
\begin{equation}
M^T_{V_1V_2} = \left[\left(\sum_i E_i\right)^2-\left(\sum_i p_{z,i}\right)^2\right]^{1/2}\,,
\end{equation}
where the index $i$ runs over all the leptons in the final state,
and assuming that the neutrinos longitudinal momenta are zero~\cite{Sirunyan:2020gyx}.

In order to provide a handle on possible issues concerning the validity of the ALP EFT expansion~\cite{Brivio:2021fog} and to estimate the impact of the highest-energy bins, we introduce an upper cut on $M_{V_1 V_2}$, that is applied on the signal simulation only.
We consider two benchmark selections: $M_{V_1V_2} <\unit[2]{TeV}$ and $M_{V_1V_2} <\unit[4]{TeV}$. These cuts are satisfied, respectively, by $85\%$ and $>99\%$ of the events in the ALP generated samples and mainly impact the signal predictions in the last bin of each distribution, as shown in Figs.~\ref{fig:dist_ZZ}--\ref{fig:dist_WW}.

The log-likelihood is constructed based on a Poisson distribution. For each VBS channel, it has the form:
\begin{align}\label{eq.likelihood}
\log L(\cB,\cW) &=\sum_k \Bigg[- \left( B_k + S_k(\cB,\cW) \right) + D_k \log \left(B_k+S_k(\cB,\cW)\right)\Bigg]
\end{align}
where the index $k$ runs over the bins of the relevant distribution. 
The number of events for the data ($D_k$) and for the SM background predictions $(B_k$) are taken from the CMS experimental publications. 
The expected number of signal events ($S_k$) accounts for both the pure ALP EW VBS signal and the ALP-SM interference contributions, that are parameterised as fourth- and second-degree polynomials in $(\cW/f_a,\ \cB/f_a)$ respectively, as explained in Sec.~\ref{sec:VBS}. The combined log-likelihood is simply constructed as the sum of $\log L$ for the individual channels.

Systematic uncertainties affecting the SM background distributions are considered fully correlated among bins of a distribution, but uncorrelated among different VBS channels. 
They are described by one nuisance parameter for each VBS channel, that multiplies both background and ALP signal yields, and is taken to be Gaussian-distributed. 
The systematic uncertainty on the signal prediction is implemented analogously and applied to $S_k$ only. It is taken to be fully correlated across all channels and bins and we assign it a total size of 20\%, obtained adding in quadrature the 16\% uncertainty on the signal normalization, the 11\% uncertainty on the renormalization and factorization scales choice and the 4\% error related to the PDFs.

\section{Results}\label{sec:results}

\subsection{Results from LHC Run~2 Measurements}
\begin{table}[t]
    \centering
    \renewcommand{\arraystretch}{1.3}
    \begin{tabular}{l|*5{>{$}c<{$}}}
    \toprule
    Analysis &  ZZ& Z\g& W^\pm \g& W^\pm Z& W^\pm W^\pm
    \\\midrule
    Branching fraction& 0.45\%& 6.7\%& 22\% & 1.5\%& 4.8\%
    \\
    Efficiency& 35.7\%& 14.0\%& 1.6\%& 11.3\% &17.0\%
    \\\bottomrule 
\end{tabular}
\caption{Summary of branching fractions and selection efficiencies for each VBS channel. The efficiencies are relative to the simulated events in which the W and Z bosons decay to electrons or muons. 
}
\label{tab:efficiencies}
\end{table}

Table~\ref{tab:efficiencies} shows the branching fractions and selection efficiencies for each VBS channel. The latter are relative to the simulated events in which the bosons are decayed to electrons and muons. 
The products of efficiencies and branching fractions range from 0.2\% to 0.9\%. 

Results are extracted from a maximum likelihood fit of signal and background to the diboson invariant mass ($ZZ$, $Z\gamma$ and $W^{\pm}\gamma$) or transverse mass ($W^{\pm}Z$ and $W^{\pm}W^{\pm}$) distributions, individually and simultaneously in all the experimental channels used in the analysis.
The likelihood is defined as described in the previous section and the background-only hypothesis is tested against the combined background and signal hypothesis.

No significant excess was observed by CMS with respect to the SM expectations. ALP couplings $\cW/f_a$ and $\cB/f_a$ are considered excluded at 95\% confidence level (CL) when the negative log likelihood (NLL) $(-\log L)$ of the combined signal and background hypothesis exceeds 3.84/2 units the NLL of the background-only hypothesis.

\begin{figure}[t]
\begin{center}
\includegraphics[width=12cm]{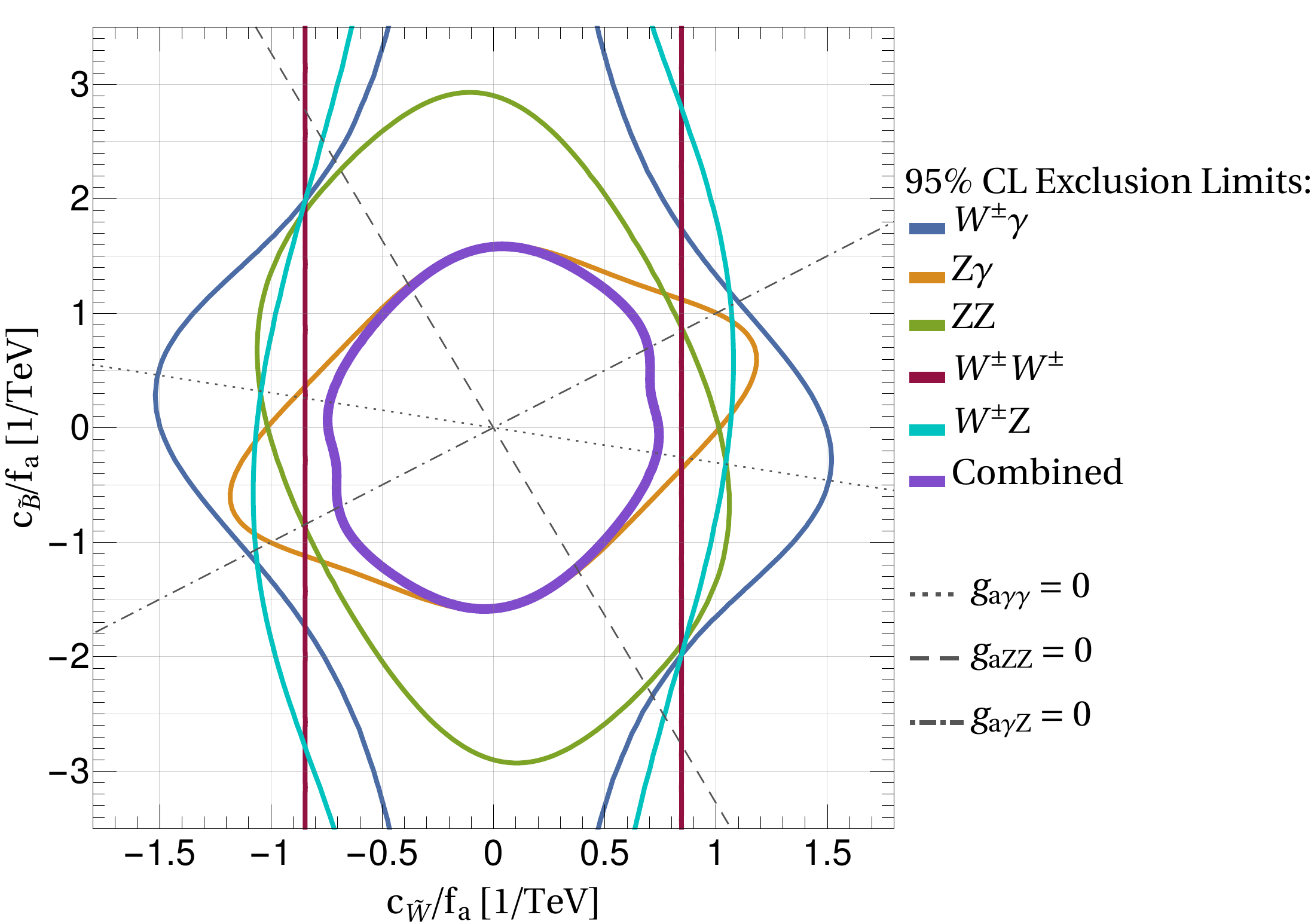}
\caption{Observed 95\% CL exclusion limits in the $(\cW/f_a,\ \cB/f_a)$ plane using the data of the Run~2 CMS publications and signal events with $M_{V_1 V_2} < \unit[4]{TeV}$. The limits have been calculated individually for the five different experimental channels considered and for their combination.
The thin dotted, dashed and dot-dashed lines indicate the directions of vanishing couplings to neutral gauge bosons.}
\label{fig:bounds_current}
\end{center}
\end{figure}

Fig.~\ref{fig:bounds_current} shows the observed 95\%~CL exclusion limits in the $(\cW/f_a, \cB/f_a)$ plane using the data of the Run~2 CMS publications and signal events with $M_{V_1 V_2}<\unit[4]{TeV}$. The limits have been calculated individually for the five different experimental channels considered and for their combination.
Tab.~\ref{tab:bounds} reports the upper bounds obtained projecting the combined 95\%~CL allowed region  onto different directions in the $(\cW/f_a, \cB/f_a)$ plane, namely the two axes and the combinations corresponding to the ALP couplings to physical bosons defined in Eq.~\eqref{eq.Lalp_physical}, which are orthogonal to the dotted, dashed and dot-dashed lines in Fig.~\ref{fig:bounds_current}. Tab.~\ref{tab:bounds} also presents the 95\%~CL limits obtained with the more conservative cut $M_{V_1V_2}<\unit[2]{TeV}$, which are about 10--15\% weaker than the ones in Fig.~\ref{fig:bounds_current}. The modest impact of this additional cut indicates that the ALP VBS cross section does not grow indefinitely with energy (see also Fig.~\ref{fig:variables}). Instead, only a small number of signal events populate the very high $M_{V_1 V_2}$ region.

In most of the parameter space, the limits are dominated by the $Z\gamma$ measurement, that is the most stringent along the $\cB$ direction. The only other measurement capable of bounding this parameter is $ZZ$, which however pays the price of the small ${\rm Br}(Z\to\ell\ell)$ and the current loose selection cuts on the dijet system. The sensitivity of the $W^\pm \g$ channel is reduced by the smaller integrated luminosity of the published CMS analysis. 
A measurement of the $\gamma\gamma$ VBS final state at large diphoton invariant masses, that has not been performed by ATLAS or CMS to date, would bring additional sensitivity to $\cB$, with a great potential for improving the current bounds~\cite{Florez:2021zoo}.

\begin{table}[t]\centering
\renewcommand{\arraystretch}{1.3}
\begin{tabular}{l|*2c|*2c|*2c}
\toprule
Coupling  & 
\multicolumn{2}{c|}{Run~2 Observed (Expected)} & 
\multicolumn{2}{c|}{$\unit[300] {fb^{-1}}$} &  
\multicolumn{2}{c}{$\unit[3000] {fb^{-1}}$} \\
\cline{2-7}
[$\unit{TeV^{-1}}$]& $M_{V_1 V_2}<\unit[4]{TeV}$ & $<\unit[2]{TeV}$
& $<\unit[4]{TeV}$ & $<\unit[2]{TeV}$
& $<\unit[4]{TeV}$ & $<\unit[2]{TeV}$\\
\midrule
$|\cW / f_a|$ & 
0.75 (0.83) & 0.86 (0.94) & 
0.71 & 0.80 &
0.55 & 0.62\\
$|\cB / f_a|$ & 
1.59 (1.35) & 1.73 (1.47) &
1.12 &  1.23 &
0.79 & 0.87\\
\midrule
$|g_{a\g\g}|$ & 
4.99 (4.24) & 5.45 (4.63) &
3.50 & 3.84 &
2.43 & 2.68 \\
$|g_{a\g Z}|$ & 
5.54 (4.74) & 6.15 (5.25) &
3.98 & 4.42 &
2.94 & 3.30 \\
$|g_{a Z Z}|$ & 
2.84 (3.02) & 3.19 (3.38) &
2.53 & 2.81 &
1.94 & 2.16 \\
$|g_{a W W}|$ & 
2.98 (3.33) & 3.43 (3.74) &
2.84 & 3.18 &
2.21 & 2.49 \\
\bottomrule
\end{tabular}
\caption{95\% CL upper limits on the absolute value of the Wilson coefficients $\cW/f_a$ and $\cB/f_a$ and projected onto the ALP couplings to physical bosons, Eq.~\eqref{eq.Lalp_physical}. The various columns report current bounds extracted from CMS Run~2 measurements and projected sensitivities for $\sqrt{s} = \unit[14]{TeV}$ and LHC higher luminosities, for signal events with $M_{V_1 V_2}$ below 4 TeV or 2 TeV.}\label{tab:bounds}
\end{table}

\subsection{Prospects for LHC Run~3 and HL-LHC}
\begin{figure}[t]
\begin{center}
\includegraphics[width=14cm]{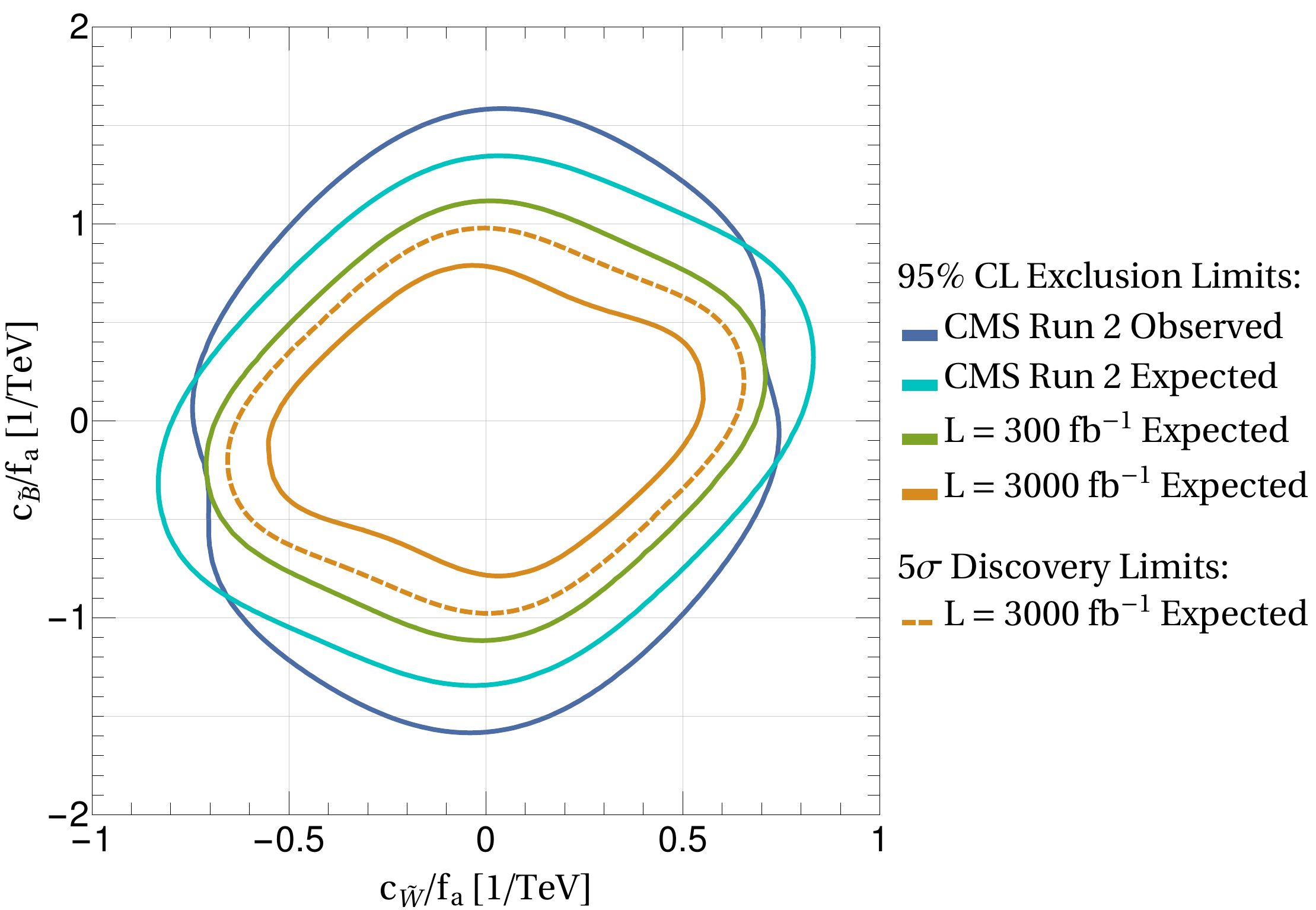}
\caption{Projected 95\% CL upper limits on the couplings $(\cW/f_a, \cB/f_a)$ for $\sqrt{s} = \unit[14]{TeV}$, $M_{V_1 V_2}<\unit[4]{TeV}$ and integrated luminosities of $300$ (green) and $\unit[3000]{fb^{-1}}$ (orange), obtained combining all VBS channels.
The blue and light blue lines show, for comparison, the observed and expected limits with Run~2 luminosities.
The dashed orange line marks the $5\sigma$-discovery limit for the HL-LHC. }
\label{fig:bounds_projected}
\end{center}
\end{figure}

In this Section we investigate the sensitivity of the nonresonant ALP VBS searches at the LHC Run~3 and HL-LHC. For simplicity, we apply the same selection criteria as the CMS Run~2 analyses, and rescale the integrated luminosities to $\unit[300]{fb^{-1}}$ and $\unit[3000]{fb^{-1}}$, respectively. An additional scaling factor $\kappa$ is applied to account for an increase in the proton collision center-of-mass energy from 13 to 14~TeV. In our approximation, $\kappa$ is taken to be constant over all distribution bins and identical for all VBS channels. Using \MGMCatNLO\ and the cuts in Eq.~\eqref{eq.gen_cuts}, we obtain $\kappa$-factors of $1.14$, $1.26$ and $1.20$ for the SM background, the ALP EW VBS signal and their interference, respectively.

Fig.~\ref{fig:bounds_projected} shows the projected 95\% CL upper limits in the $(\cW/f_a, \cB/f_a)$ plane
for  $\sqrt{s} = \unit[14]{TeV}$, $M_{V_1 V_2} < \unit[4]{TeV}$ and integrated luminosities $\unit[300]{fb^{-1}}$ and $\unit[3000]{fb^{-1}}$. For comparison, the observed and expected Run~2 limits have been included as well. 
The interplay between the individual channels is not shown in Fig.~\ref{fig:bounds_projected} as it remains qualitatively unchanged compared to Fig.~\ref{fig:bounds_current}. As expected, the largest individual improvement is found for the $W^\pm \g$ channel. However, the combined limits are still dominated by the $Z\g$ channel and with a significant contribution of $W^\pm W^\pm$ for the highest values of $\cW/f_a$. 
We find that the bounds on $\cB$ can improve by roughly a factor 2 at the HL-LHC compared to current constraints, while those on $\cW$ by a factor $\sim 1.4$. 

Fig.~\ref{fig:bounds_projected} also shows, for reference, the curve corresponding to the expected discovery limit for $\sqrt{s} = \unit[14]{TeV}$, $M_{V_1 V_2} < \unit[4]{TeV}$ and an integrated luminosity of $\unit[3000]{fb^{-1}}$, defined as the set of $(\cW/f_a,\cB/f_a)$ values for which the SM point is excluded by 5 standard deviations, assuming that the measurement matches the predicted ALP EW VBS signal.
The fact that it is fully contained inside the projected exclusion limits for current and Run~3 luminosities indicates that null results at previous LHC Runs will not exclude a priori the possibility of a discovery at the HL-LHC.

\subsection{Dependence on the ALP Mass and Decay Width}
\label{sec:mass_width}
\begin{figure}[t]
\centering 
\includegraphics[width=12cm]{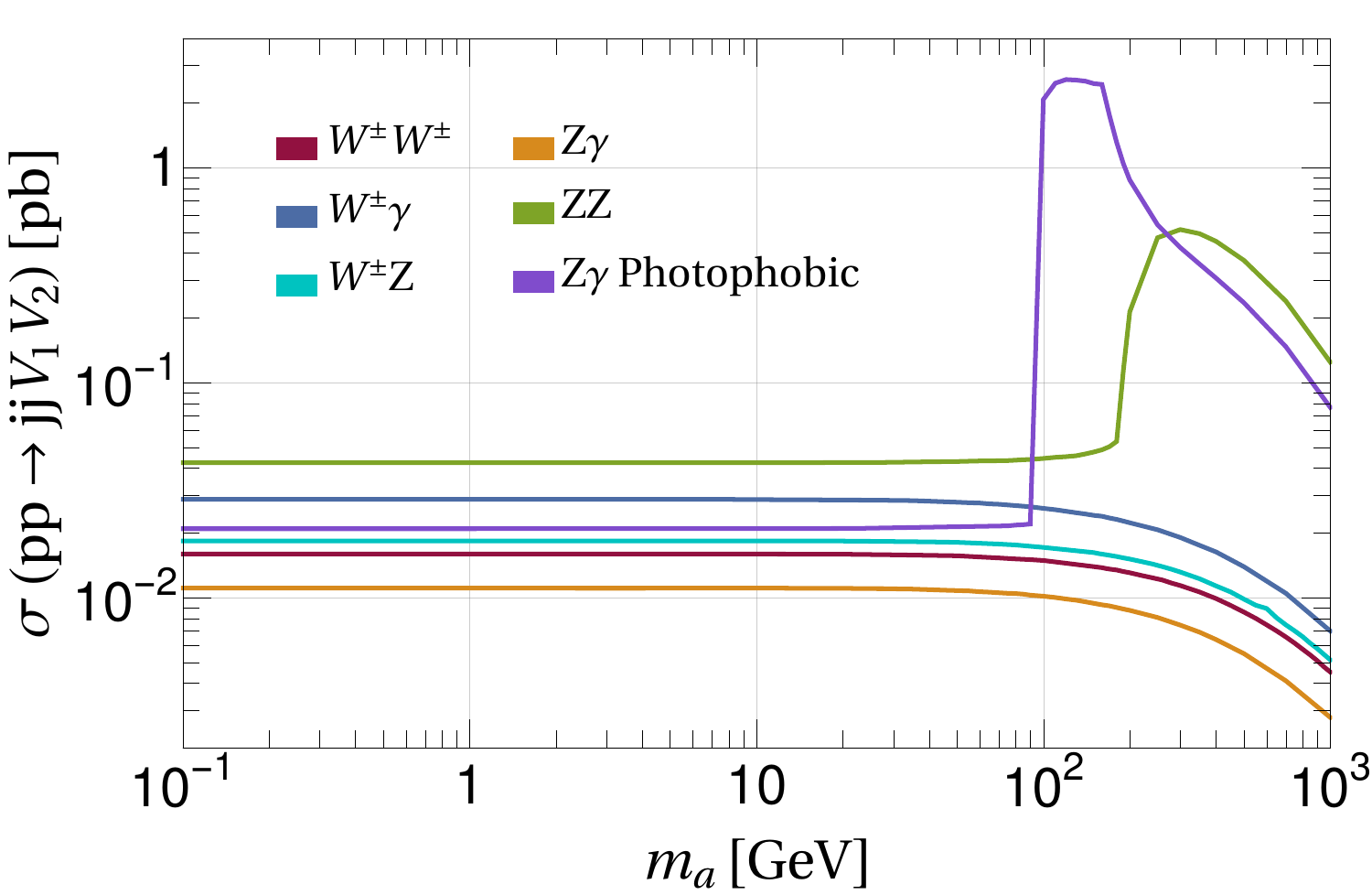}
\caption{Total cross sections at $\sqrt{s} = \unit[13]{TeV}$ for the ALP contributions to the different VBS channels as a function of the ALP mass. All lines are evaluated at $\cW/f_a=\cB/f_a=\unit[1]{TeV^{-1}}$, that corresponds to the benchmark point $p_0$ in Eq.~\eqref{eq.points}. The exception is the "$Z\g$ photophobic" case, that is evaluated at $p_4$ instead. At each point in the plot, the ALP decay width was re-computed as a function of $m_a,\cW$ and $\cB$. }
\label{fig.mass}
\end{figure}
 
Our results were derived assuming that the ALP gives only off-shell contributions to all VBS processes considered. Specifically, in the simulations we fixed the ALP mass and decay width to $m_a=\unit[1]{MeV}, \Gamma_a=0$, which satisfy $\sqrt{|p_a^2|} \gg m_a,\, \Gamma_a$, being $p_a$ the momentum flowing through the ALP propagator. As long as this kinematic condition is verified, the bounds are essentially independent of the specific $m_a$ and $\Gamma_a$ assumed. This is an important difference with respect to resonant searches, that only apply for limited mass and width windows.

Figure~\ref{fig.mass} provides a basic check of the validity of the off-shell approximation, showing the cross section for the ALP signal at $\sqrt{s} = \unit[13]{TeV}$ with the cuts in Eq.~\ref{eq.gen_cuts}, as a function of $m_a$ for fixed values of $\cW$, $\cB$ and $f_a$. The width $\Gamma_a$ was implicitly computed at every point as a function of $m_a$ and of the ALP couplings, and it scales as $\Gamma_a\propto m_a^3(c_i/f_a)^2$. The lines in Fig.~\ref{fig.mass} extend indefinitely to the left, confirming that the simulations apply to arbitrarily small $m_a$. 
In the direction of larger $m_a$ the cross sections for $W^\pm Z, W^\pm\gamma, W^\pm W^\pm$ start falling once the $t$-channel propagator becomes kinematically dominated by the ALP mass.
For the $Z\gamma$ and $ZZ$ channels, the resonant behavior is visible for $(\cW,\cB)$ benchmark points that allow the ALP exchange in $s$-channel.
As $g_{a\g Z}=0$ is enforced at $p_0$, we evaluate the $Z\gamma$ channel also at the "photophobic" point $p_4$ in order to test the resonant case. 

Based on these indications, our results can be safely taken to hold up to $m_a\lesssim\unit[100]{GeV}$. At this mass, the $ZZ$ and $W^\pm V$ cross sections have deviated by about 10\% from their asymptotic values for $m_a\to 0$. At the same time, the $Z\gamma$ resonance is present but not visible in the CMS measurement, that requires $M_{Z\g}>\unit[160]{GeV}$~\cite{CMS:2021gme}.

\section{Comparison to Existing Bounds}\label{sec:comparison}
\begin{figure}[t]
\centering
\includegraphics[width=.495\textwidth]{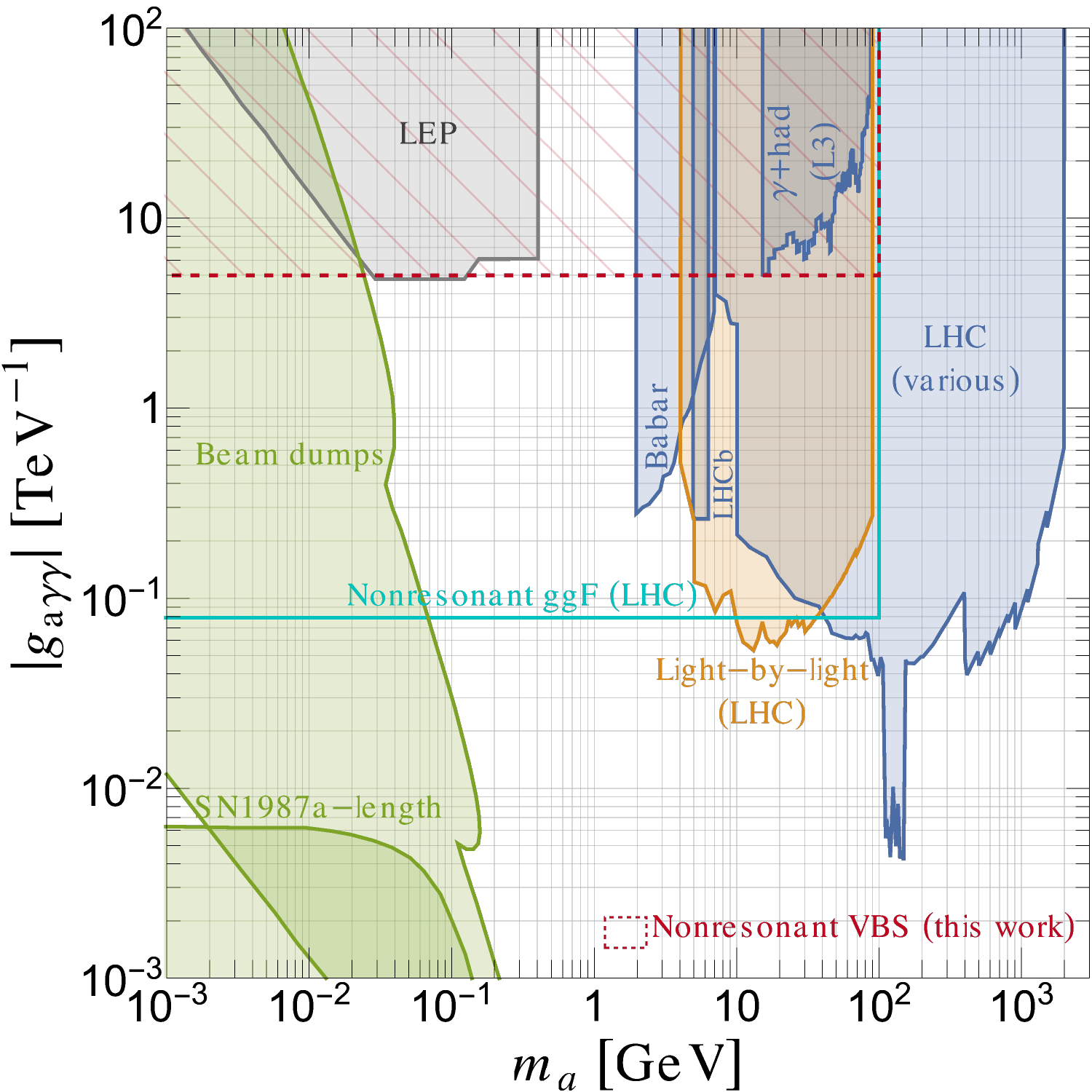}
\includegraphics[width=.495\textwidth]{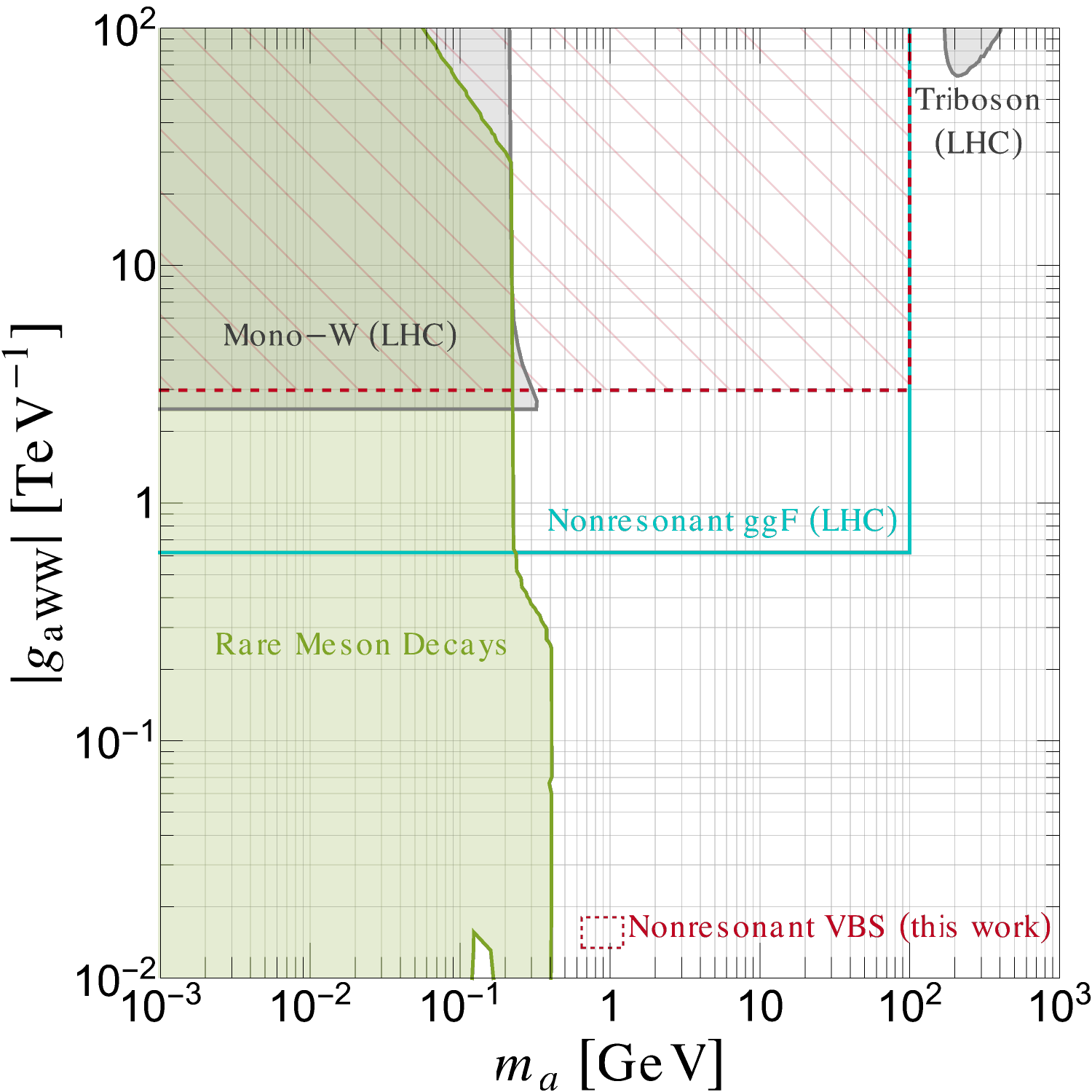}
\\
\includegraphics[width=.495\textwidth]{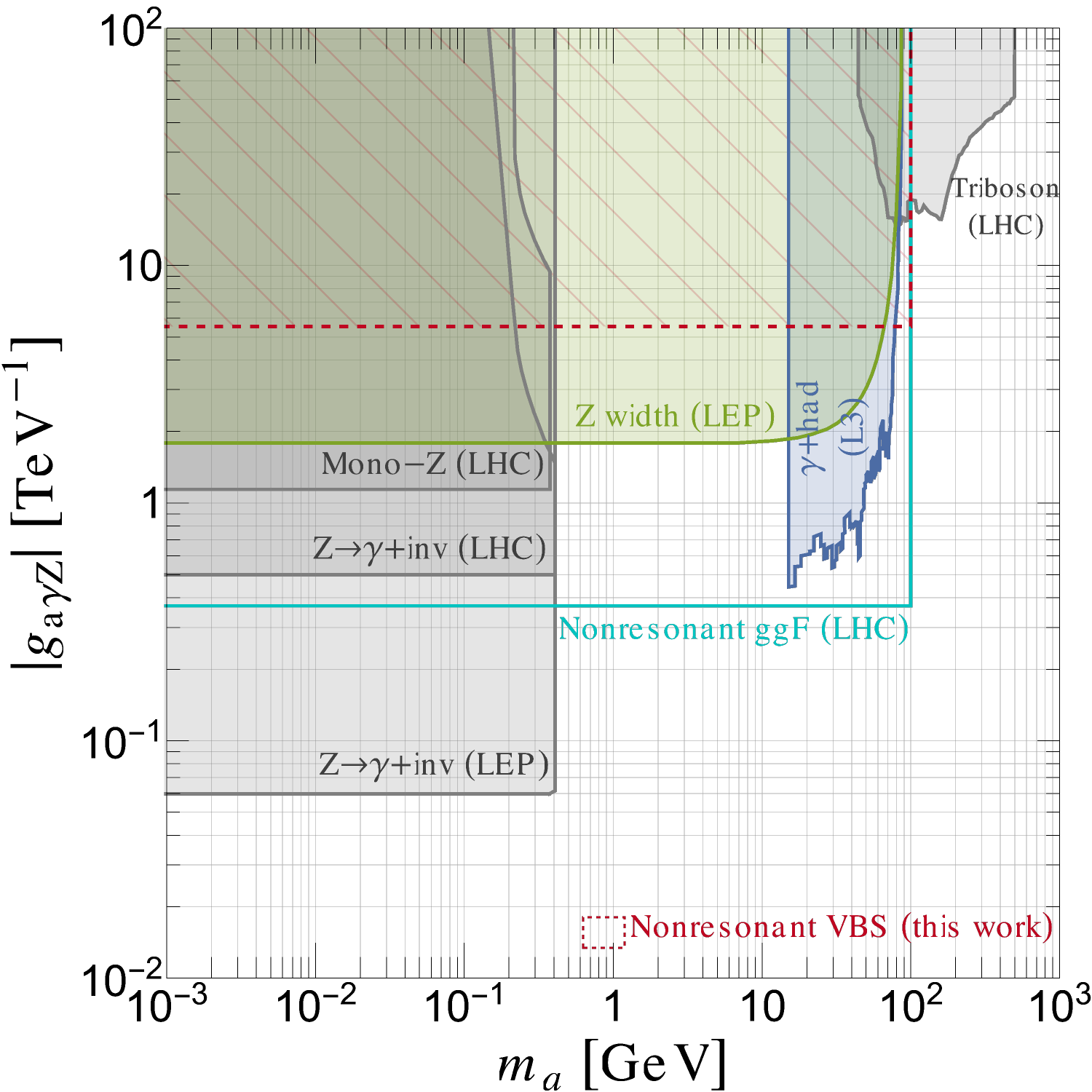}
\includegraphics[width=.495\textwidth]{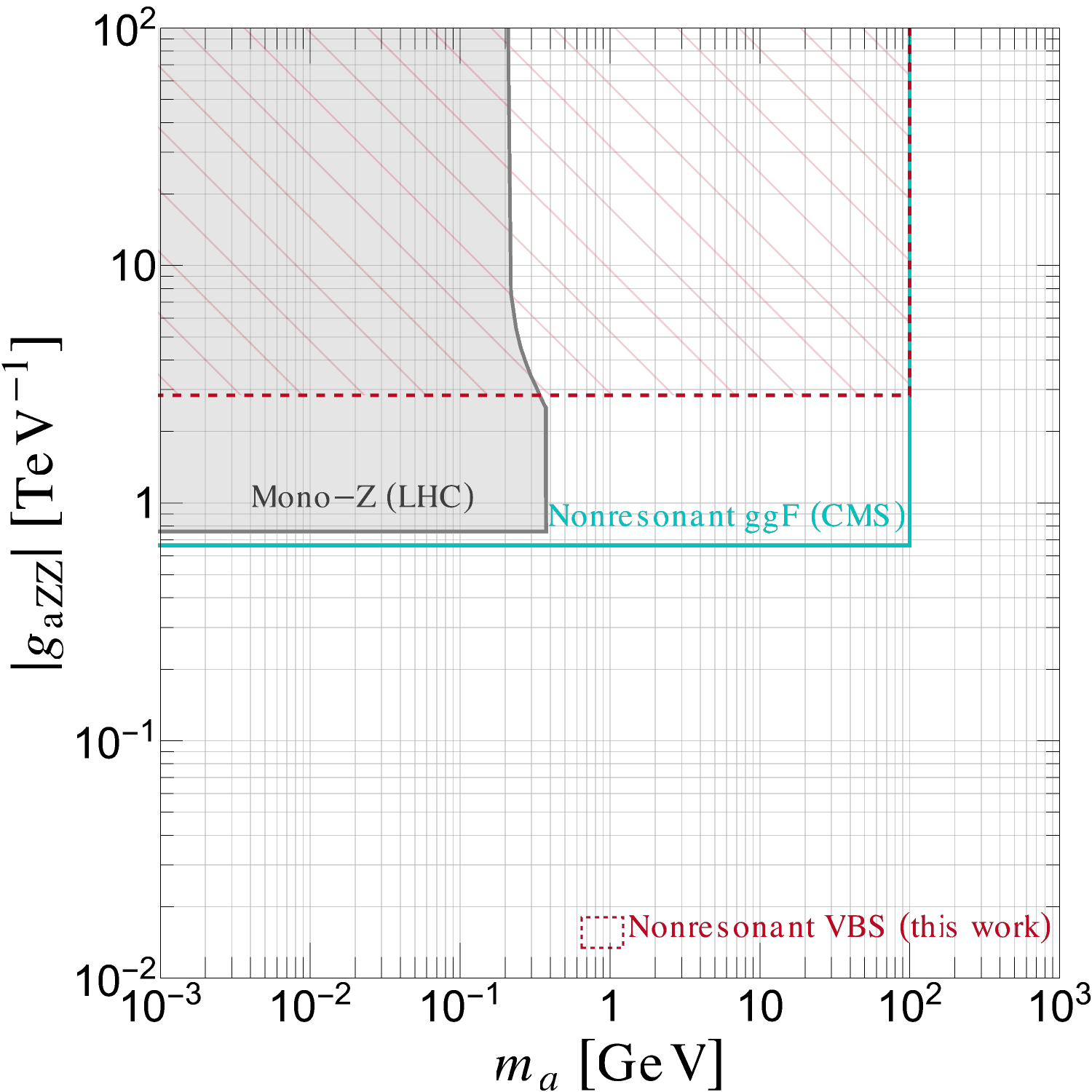}
\caption{Summary of current constraints on ALP couplings to EW gauge bosons defined in Eq.~\eqref{eq.Lalp_physical}, as a function of the ALP mass $m_a$. Limits derived in this work are labeled "Nonresonant VBS" and shown in {\color{BrickRed}\bf red}. Previous constraints are shown with a color coding that indicates different underlying theory assumptions. 
{\color{BurntOrange}\bf Orange} indicates a $\Br(a\to\g\g)=1$ assumption, {\color{Blue}\bf dark blue} indicates an assumed gluon dominance $g_{agg}\gg g_{aV_1V_2}$, while bounds in {\color{Turquoise}\bf light blue} scale with $1/g_{agg}$ and are given for $g_{agg}=\unit[1]{TeV^{-1}}$.
{\color{darkgray}\bf Grey} indicates more complex assumptions on the ALP EW couplings. Genuine bounds, that hold without further assumptions, are in {\color{OliveGreen}\bf green}.
See the main text for more details.
}
\label{fig.comparison}
\end{figure}

Fig.~\ref{fig.comparison} shows the observed bounds obtained in this work as a function of the EW $g_{aVV}$ couplings defined in Eq.~\eqref{eq.Lalp_physical}, and of $m_a$, compared to previously derived bounds.  The numerical results of our study are also reported in Tab.~\ref{tab:bounds} for observed, expected and projected limits.

Most of the constraints shown in the Figure are taken from the compilation in Ref.~\cite{Alonso-Alvarez:2018irt} and updated to include more recent results. For ALP masses in the MeV-GeV window and within the range shown, the ALP coupling to photons is constrained by beam-dump experiments~\cite{Dobrich:2015jyk,Bjorken:1988as,Riordan:1987aw,Blumlein:2013cua}, by new physics searches in $e^+e^-\to 2\g,3\g$ at LEP~\cite{Jaeckel:2015jla,Mimasu:2014nea} and by explosion energy arguments in supernovae~\cite{Caputo:2021rux,Caputo:2022mah} (labeled "SN").
At higher ALP masses, all constraints on $g_{a\g\g}$ are due to searches at colliders, where the ALP decays resonantly either to hadrons or to photon pairs. In the first case, the relevant processes are $\Upsilon \to\g+\text{hadrons}$ at 
BaBar~\cite{BaBar:2011kau} and $e^+e^-\to\gamma+\text{hadrons}$ at L3~\cite{L3:1992kcg}, that also constrains $g_{a\g Z}$.
In the second case, the leading bounds stem from photon pair production at the LHC, both in proton-proton collisions~\cite{Jaeckel:2012yz,Mariotti:2017vtv} (labeled "LHC" for those from ATLAS and CMS measurements and "LHCb" for those from LHCb searches~\cite{CidVidal:2018blh}) and in light-by-light scattering $\g\g\to a\to \g\g$ measured in Pb-Pb collisions~\cite{CMS:2018erd,ATLAS:2020hii} (labeled "Light-by-light (LHC)").
Most constraints on the couplings of the ALP to massive gauge bosons assume a stable ALP and cover the sub-GeV mass region. In this case, limits are inferred from mono-$W$ and mono-$Z$~\cite{Brivio:2017ije} at the LHC and, for $g_{a\gamma Z}$, from the non-observation of exotic $Z\to \gamma+\text{invisible}$ decays at LEP~\cite{Craig:2018kne} and at the LHC~\cite{ATLAS:2020uiq} (labeled "$Z\to\gamma+{\rm inv.}$ (LHC)"). If the assumption of a stable ALP is relaxed, the latter constraint can be replaced by the more conservative bound due to the measurement of the total $Z$ decay width at LEP, that extends up to $m_a\lesssim m_Z$~\cite{Brivio:2017ije,Craig:2018kne}. In the region where the ALP can decay to hadrons, the same process leads to $Z\to\gamma+\text{hadrons}$~\cite{L3:1992kcg}.  
The ALP coupling to W bosons is the only one contributing to rare meson decays at 1-loop, which allow to set very stringent limits for light ALPs~\cite{Izaguirre:2016dfi,BNL-E949:2009dza}. 
For ALP masses above 100 GeV, the dominant bounds stem from resonant triboson searches~\cite{Craig:2018kne}.
Finally, nonresonant searches in diboson production via gluon fusion at the LHC (labeled "Nonresonant ggF") allow to constrain all four ALP interactions. Each nonresonant bound is extracted from a single process $gg\to a^*\to V_1 V_2$: the constraint on $g_{a\g\g}$ was derived in Ref.~\cite{Gavela:2019cmq}, those on $g_{aWW},g_{a\g Z}$ in Ref.~\cite{Carra:2021ycg}, and the constraint on $g_{aZZ}$ in Ref.~\cite{CMS:2021xor}.

An important aspect to consider is that, in general, any given measurement can depend on several ALP couplings. In order to represent the corresponding bound in the 2D ($m_a,g_{aVV}$) plane, it is then necessary to define a projection rationale or introduce theoretical assumptions, which can vary significantly from constraint to constraint. These differences should be taken into account for a proper comparison.
In Fig.~\ref{fig.comparison}, the bounds derived in this work (red dashed)
are those corresponding to the 95\% C.L. limits in Tab.~\ref{tab:bounds}. As they are derived from the allowed region in the $(\cW/f_a,\cB/f_a)$ plane,  they automatically take into account gauge invariance relations. Because of the arguments laid down in Sec.~\ref{sec:Lagrangian}, they also have limited sensitivity to the coupling to gluons.
The remaining bounds are derived with alternative strategies, that we highlight with color coding in Fig.~\ref{fig.comparison}. Bounds that apply without extra assumptions, are reported in green. The bounds drawn in light blue, that include nonresonant $gg\to a^*\to V_1V_2$ processes, 
scale with $1/g_{agg}$ and for $\cG\to0$ are lifted completely. In the Figure, they are normalized to $g_{agg}=\unit[1]{TeV^{-1}}$. Bounds drawn in dark blue assume gluon-dominance, i.e. $g_{agg}\gg g_{aV_1V_2}$, and in this limit they are largely independent of $\cG$, see Ref.~\cite{Alonso-Alvarez:2018irt}. Among these, bounds on $g_{a\g\g}$ labeled as "LHC" additionally assume negligible branching fractions to fermions and heavy EW bosons in the mass region where they are kinematically allowed.  The limit from light-by-light scattering, shown in orange, assumes ${\rm Br}(a\to\g\g)=1$, which corresponds to vanishing couplings to gluons and light fermions.
Bounds that make more elaborate assumptions about the ALP parameter space or assumptions on the EW sector itself are shown in grey. Among these, triboson constraints on $g_{aWW}$ and $g_{a\g Z}$ assume a photophobic ALP scenario~\cite{Craig:2018kne}. All searches for a stable ALP (mono-$W$, mono-$Z$, $Z\to \g+{\rm inv.}$) implicitly assume a small enough ALP decay width, which, in the relevant mass range, translates into assumptions on the coupling to photons, electrons and muons.
The LEP constraints assume negligible branching fractions to leptons.
Note also that this bound is truncated to $m_a\leq 3m_\pi\simeq \unit[0.5]{GeV}$ because, beyond this threshold, hadronic ALP decay channels are kinematically open. This would introduce a further dependence on $\cG$ whose modeling would require a dedicated analysis~\cite{Alonso-Alvarez:2018irt}.
Constraints derived with assumptions that explicitly violate the gauge invariance relations, e.g. by explicitly requiring only one non-zero EW coupling, are omitted.

Overall, we find that the main value of nonresonant searches in VBS is that they probe the ALP interactions with EW bosons directly (at tree level) and independently of the coupling to gluons. 
In particular, nonresonant VBS constraints are stronger than those from nonresonant diboson production whenever $g_{agg}$ is smaller than a certain threshold, that roughly ranges between $\unit[0.01]{TeV^{-1}}$ and $\unit[0.2]{TeV^{-1}}$ depending on the EW coupling of interest.  For cases where the ALP-gluon coupling is very suppressed, such as Majorons,\footnote{A priori, the ALP-gluon interaction is not protected by any symmetry. Therefore, technically, it cannot be assumed to be exactly vanishing, even starting from a $\cG=0$ condition. In the Majoron case it is generated at 2-loops~\cite{Heeck:2019guh} and therefore remains very suppressed.} 
VBS bounds are the most stringent in the 0.5--100~GeV mass region for $g_{aWW}$, $g_{aZZ}$, and in the 0.5--4~GeV region for $g_{a\g\g}$.
In the case of $g_{a\g Z}$, the current best bounds for $m_a<m_Z$ come from the total $Z$ width measurement at LEP.

\section{Conclusions}\label{sec:conclusions}
We have investigated the possibility of constraining EW ALP interactions via the measurement of EW VBS processes at the LHC, where the ALP can induce nonresonant signals if it is too light to be produced resonantly.
We have studied the production of $ZZ$, $Z\g$, $W^\pm \g$, $W^\pm Z$ and same-sign $W^\pm W^\pm$ pairs
with large diboson invariant masses in association with two jets.
New upper limits on ALP couplings to EW bosons have been derived from a reinterpretation of Run~2 public CMS VBS analyses.
Among the channels considered, the most constraining ones are currently $Z\g$ and $W^\pm W^\pm$.

The limits have been calculated both in the plane of the gauge-invariant ALP EW couplings $(\cW/f_a,\cB/f_a)$ and projected onto the 4 mass-eigenstate couplings defined in Eq.~\eqref{eq.Lalp_physical}, to facilitate the comparison with other results.
The constraints inferred on ALP couplings to $ZZ$, $W^\pm W^\pm$ and $Z\g$ pairs are very competitive with other LHC and LEP limits for ALP
masses up to $\unit[100]{GeV}$. They probe previously unexplored regions of the parameter space and have the advantage of being independent of the ALP coupling to gluons and of the ALP decay width. 
This is important in view of a global analysis of ALP couplings, where VBS can help disentangling EW from gluon interactions.
All the constraints extracted in this work can be further improved in the future, for instance, by adopting a finer binning for the kinematic distributions, or by incorporating into the fit measurements by the ATLAS Collaboration or measurements of other VBS channels (e.g. opposite-sign $W^\pm W^\pm$ or semileptonic $ZV$). 

Simple projections for integrated luminosities up to $\unit[3000]{fb^{-1}}$ have been calculated, demonstrating the power of future dedicated analyses. 
Searches for nonresonant new physics signals in VBS production at the LHC Run~3 and HL-LHC performed by the ATLAS and CMS Collaborations will be able to probe the existence of ALPs for relevant values of their couplings to EW bosons.

\acknowledgments
We are grateful to G. Alonso and P. Quilez for providing code and information for the comparisons presented in Sec.~\ref{sec:comparison}, to E. Vitagliano for guidance on the update of supernova constraints and to B. Gavela and V. Sanz for their continuous support. We also acknowledge the contributions of J.M. No to the initial phase of this work. J.B and I.B. thank B. Heinemann, S. Heim, S. Bruggisser and P. Govoni for several illuminating discussions.
The authors acknowledge the support of the CA16108 VBSCan COST Action.
The work of J.B. was supported by the Spanish MICIU through the National Program FPU (grant number FPU18/03047). The work of J.M.R. was supported by the Spanish MICIU through the National Program FPI-Severo Ochoa (grant number PRE2019-089233). J.B. and J.M.R. acknowledge partial financial support by the Spanish MINECO through the Centro de excelencia Severo Ochoa Program under grant SEV-2016-0597, by the Spanish ``Agencia Estatal de Investigac\'ion'' (AEI) and the EU ``Fondo Europeo de Desarrollo Regional'' (FEDER) through the project PID2019-108892RB-I00/AEI/10.13039/501100011033. J.F.T. acknowledges support 
from the AEI and the EU FEDER through the project PID2020-116262RB-C43.

\appendix

\section{Expected ALP EW VBS Diboson Mass Distributions}\label{app:polynomial}
Tab.~\ref{tab:poly} reports the expected ALP EW VBS pure signal and interference cross sections at $\sqrt{s} = \unit[13]{TeV}$ as a function of the Wilson coefficients $\cW$ and $\cB$ for $f_a = \unit[1]{TeV}$, after selection cuts and $M_{V_1 V_2} < \unit[4]{TeV}$.

\begin{figure}[ht]
\centering
\includegraphics[width=12cm]{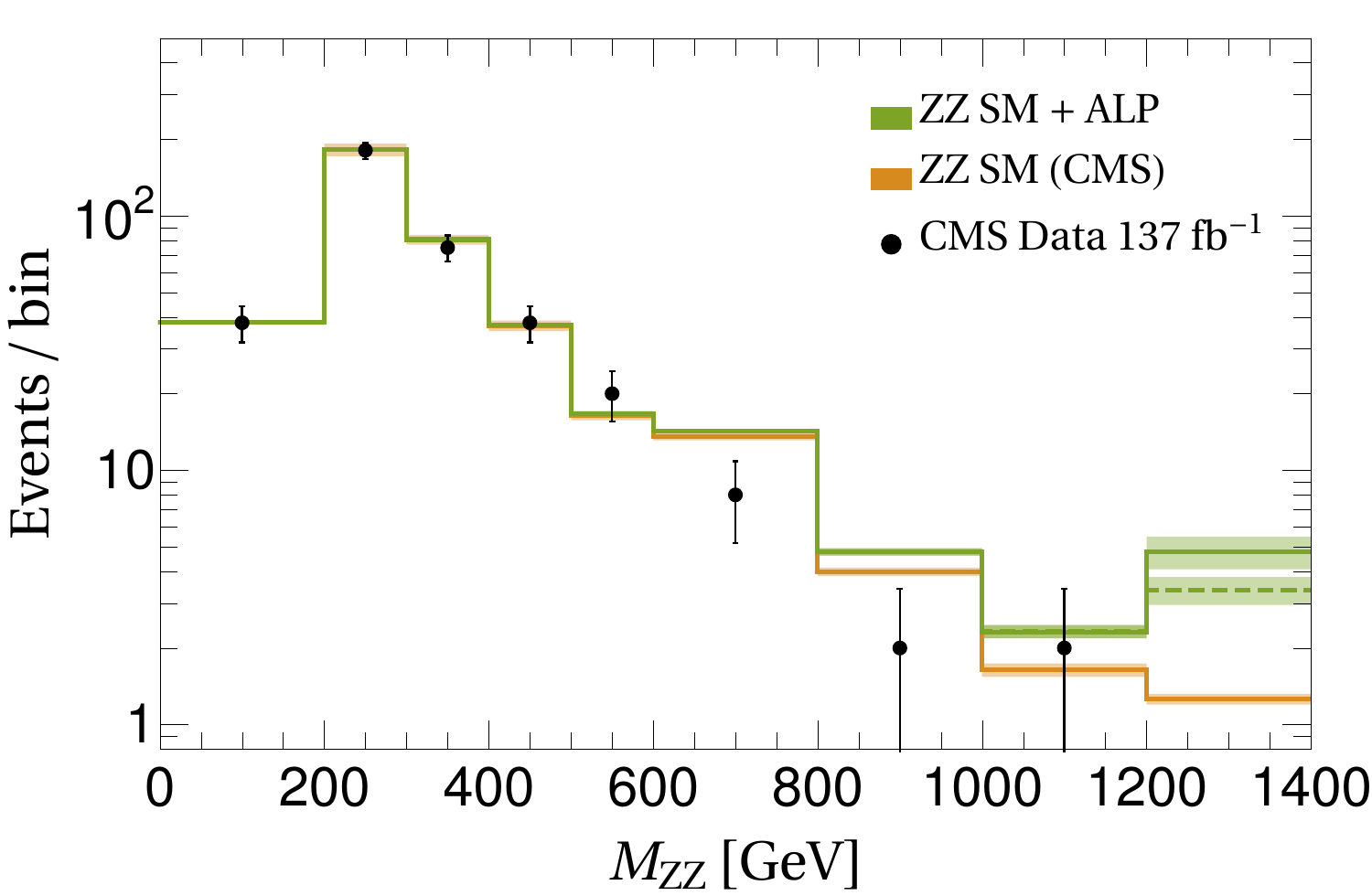}
\caption{$M_{ZZ}$ distribution for the $pp \to jjZZ \to jj\ell^+\ell^-\ell^+\ell^-$ channel. The data points and the total SM background (orange) are taken from the measurement in Ref.~\cite{Sirunyan:2020alo}. The last bin contains the overflow events. The dashed and solid green lines show the total ALP EW VBS signal for $\cB/f_a=\cW/f_a=\unit[1]{TeV^{-1}}$ with a cut of $M_{ZZ}<\unit[2]{TeV}$ and $\unit[4]{TeV}$, respectively.}
\label{fig:dist_ZZ}
\end{figure}

\begin{figure}[ht]
\centering
\includegraphics[width=12cm]{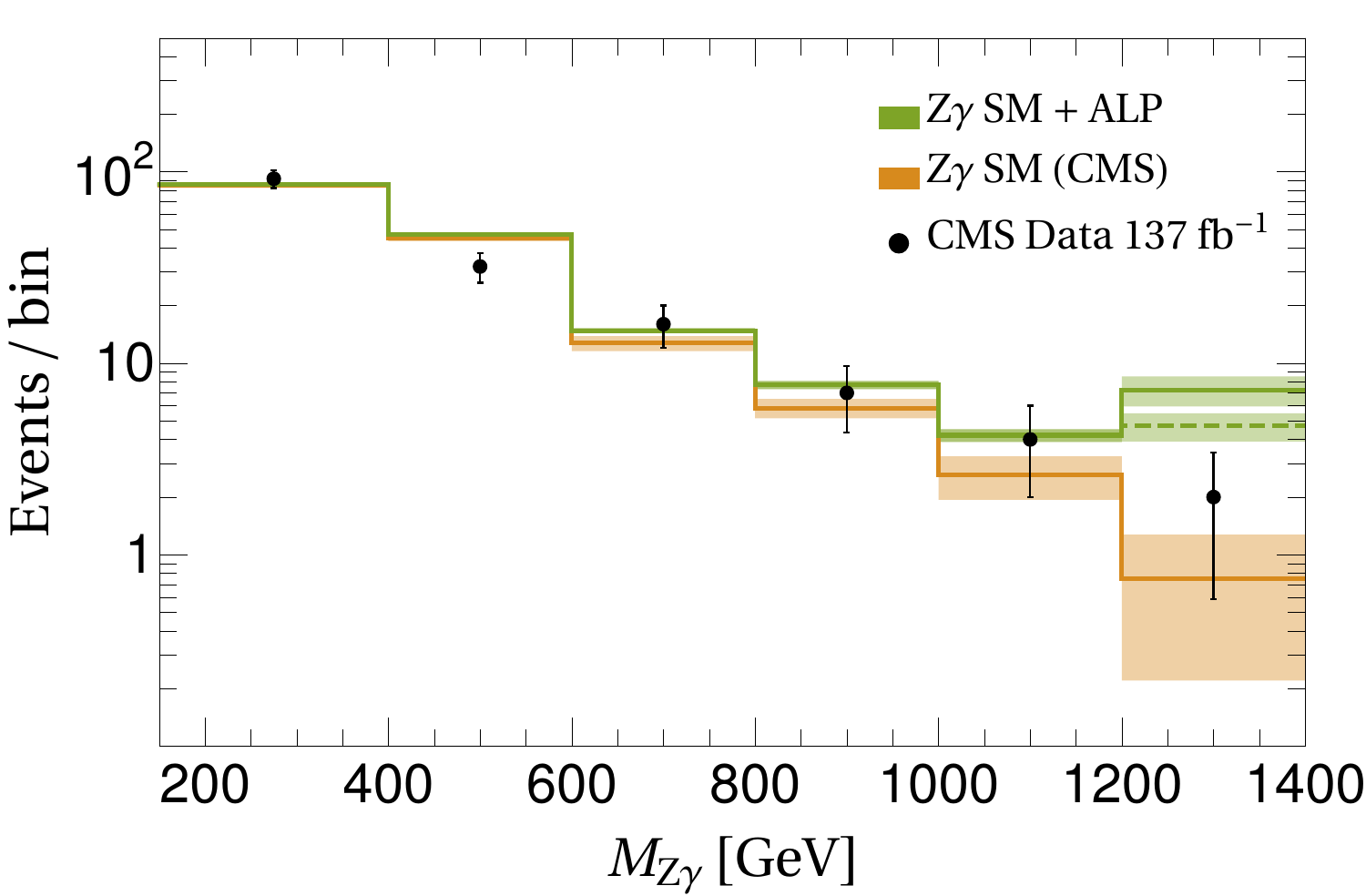}
\caption{$M_{Z\g}$ distribution for the $pp \to jjZ\g \to jj\ell^+\ell^-\g$ channel. The data points and the total SM background (orange) are taken from the measurement in Ref.~\cite{CMS:2021gme}. The last bin contains the overflow events. The dashed and solid green lines show the total ALP EW VBS signal for $\cB/f_a=\cW/f_a=\unit[1]{TeV^{-1}}$ with a cut of $M_{Z\g}<\unit[2]{TeV}$ and $\unit[4]{TeV}$, respectively.}
\label{fig:dist_Zg}
\end{figure}

\begin{figure}[ht]
\centering
\includegraphics[width=12cm]{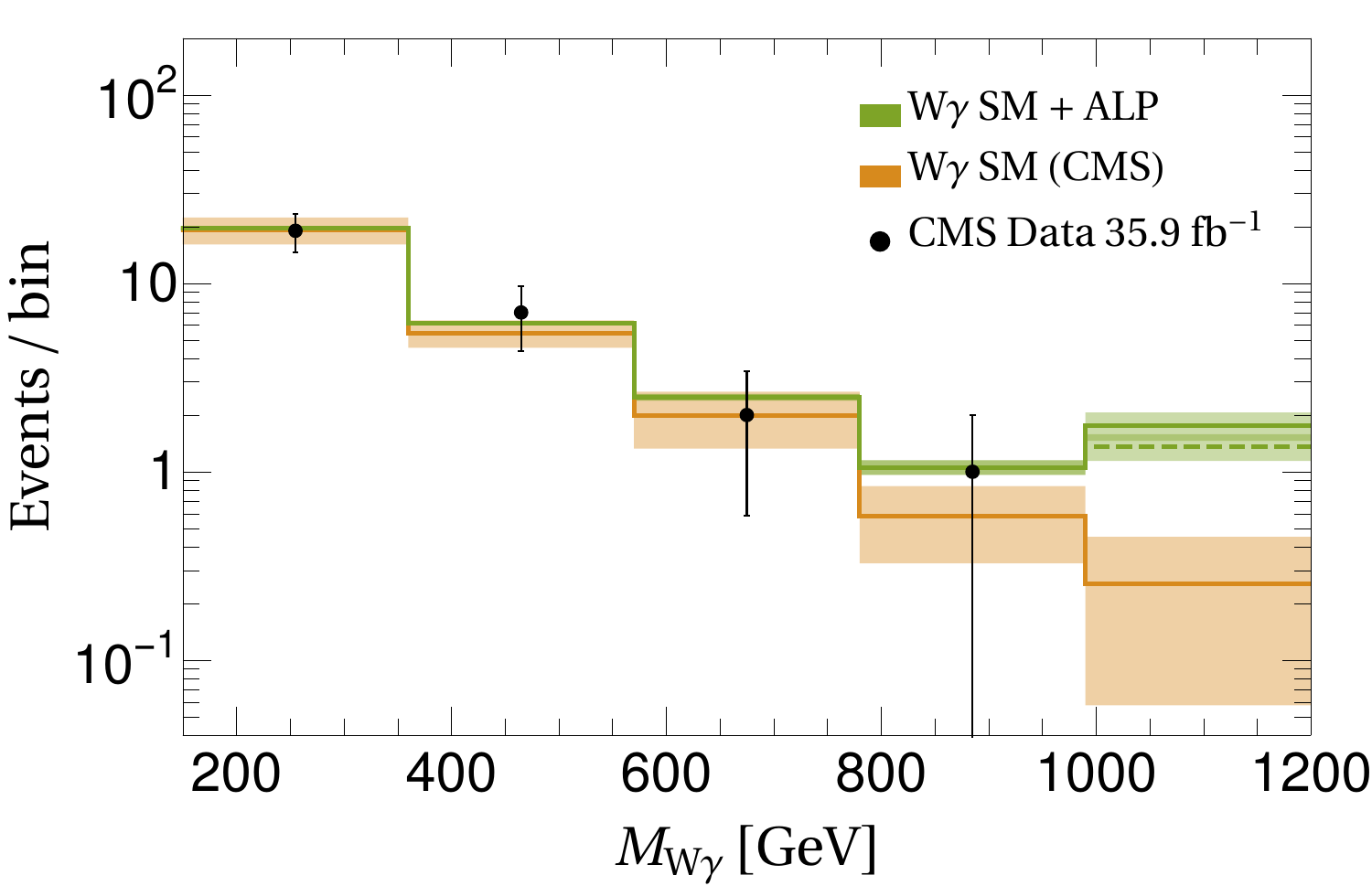}
\caption{$M_{W\g}$ distribution for the $pp \to jjW^\pm\g \to jj\g\ell^\pm\nu$ channel.The data points and the total SM background (orange) are taken from the measurement in Ref.~\cite{Sirunyan:2020azs}. The last bin contains the overflow events. The dashed and solid green lines show the total ALP EW VBS signal for $\cB/f_a=\cW/f_a=\unit[1]{TeV^{-1}}$ with a cut of $M_{W\g}<\unit[2]{TeV}$ and $\unit[4]{TeV}$, respectively.}
\label{fig:dist_Wg}
\end{figure}

\begin{figure}[ht]
\centering
\includegraphics[width=12cm]{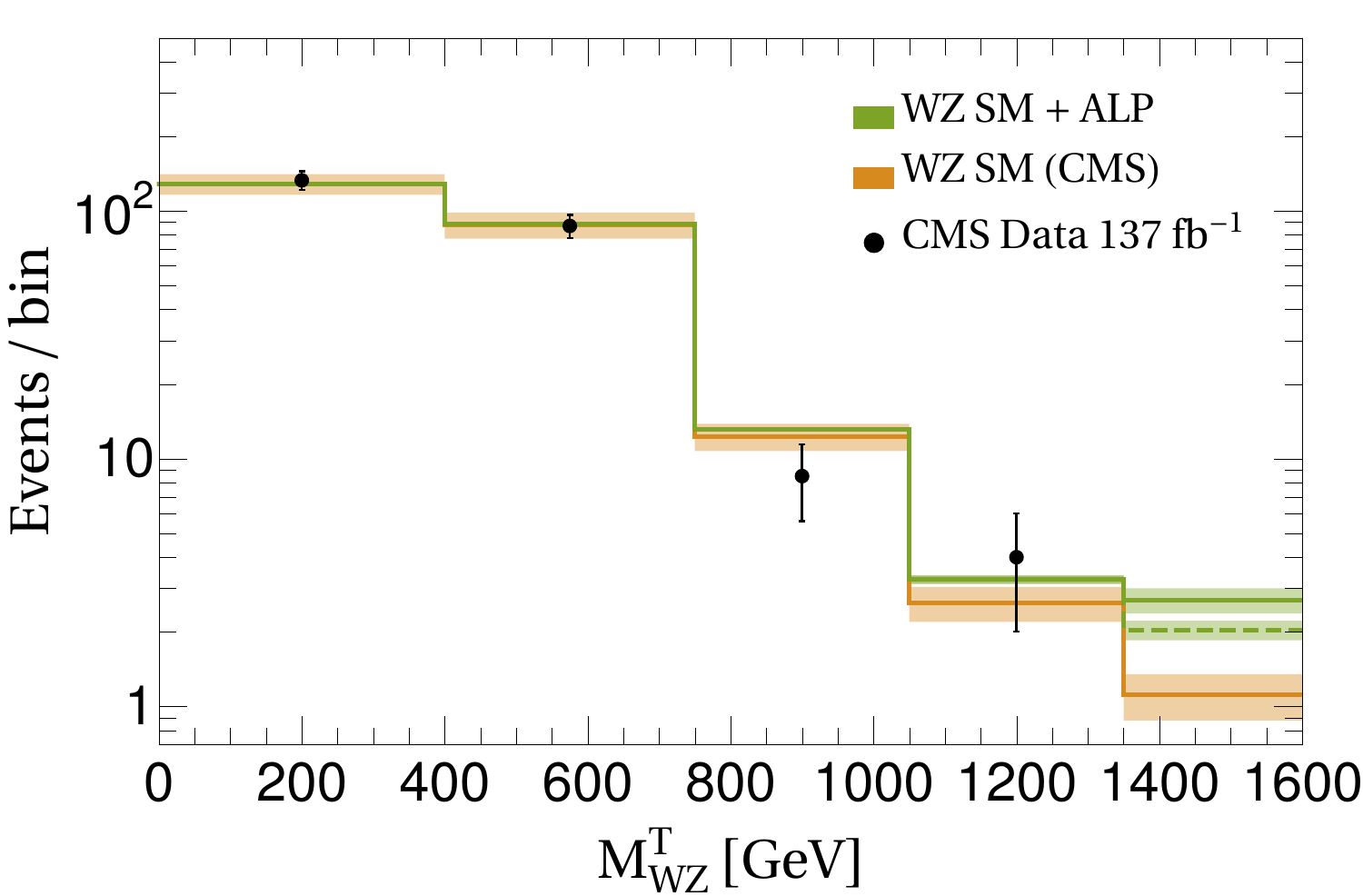}
\caption{$M^T_{WZ}$ distribution for the $pp \to jjW^\pm Z \to jj\ell^+\ell^-\ell^{\prime\pm}\nu$ channel.The data points and the total SM background (orange) are taken from the measurement in Ref.~\cite{Sirunyan:2020gyx}. The last bin contains the overflow events. The dashed and solid green lines show the total ALP EW VBS signal for $\cB/f_a=\cW/f_a=\unit[1]{TeV^{-1}}$ with a cut of $M_{WZ}<\unit[2]{TeV}$ and $\unit[4]{TeV}$, respectively.}
\label{fig:dist_WZ}
\end{figure}

\begin{figure}[ht]
\centering
\includegraphics[width=12cm]{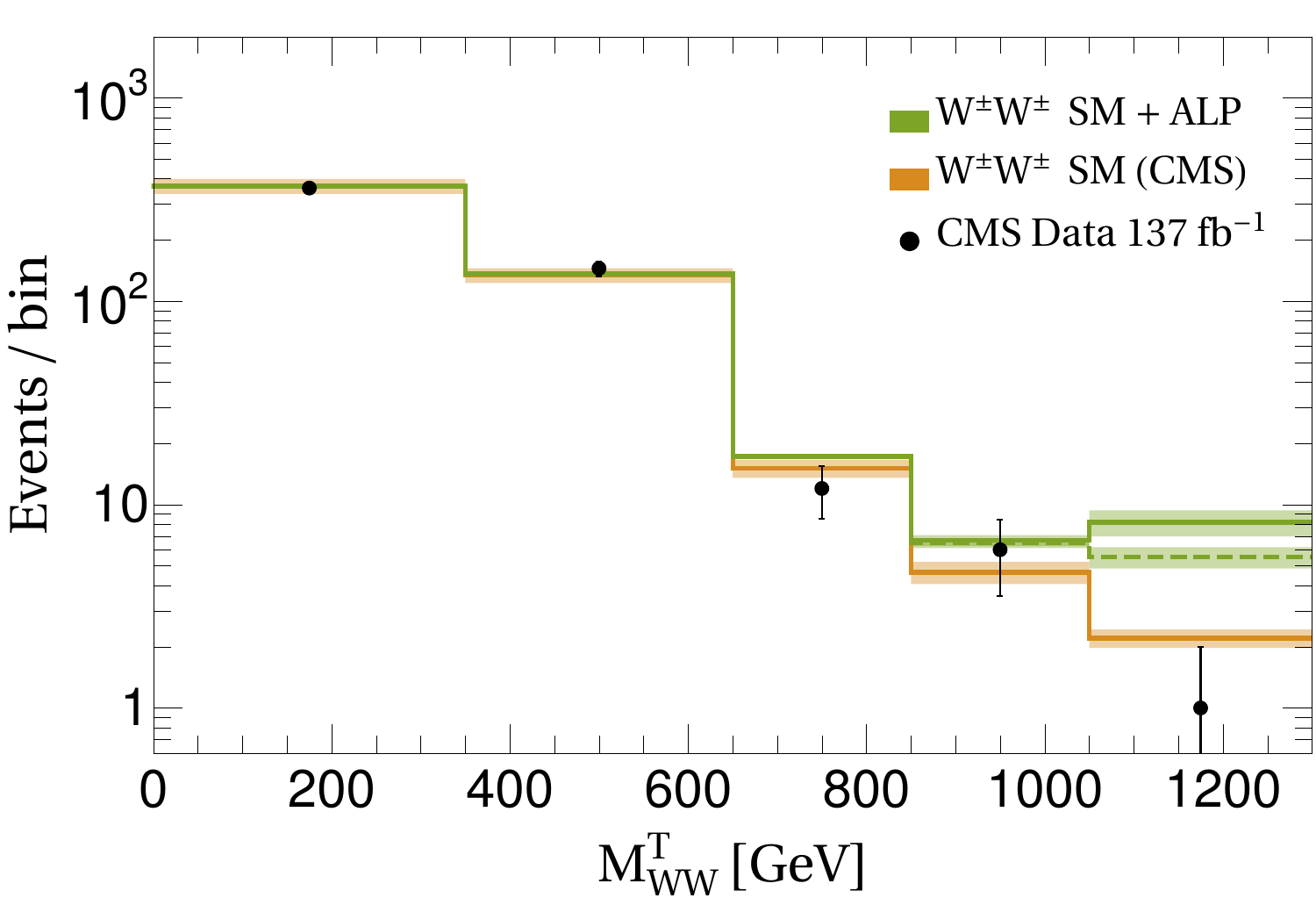}
\caption{$M^T_{WW}$ distribution for the $pp \to jjW^\pm W^\pm \to jj\ell^\pm\ell^\pm\nu\nu$ channel. The data points and the total SM background (orange) are taken from the measurement in Ref.~\cite{Sirunyan:2020gyx}. The last bin contains the overflow events. The dashed and solid green lines show the total ALP EW VBS signal for $\cB/f_a=\cW/f_a=\unit[1]{TeV^{-1}}$ with a cut of $M_{WW}<\unit[2]{TeV}$ and $\unit[4]{TeV}$, respectively.}
\label{fig:dist_WW}
\end{figure}

The diboson invariant mass or transverse mass distributions after selection cuts for the five VBS channels studied are shown in Figs.~\ref{fig:dist_ZZ}--\ref{fig:dist_WW}.
The data points and the total SM background (orange line) are taken from the CMS publications. The dashed and solid green lines represent the total ALP EW VBS signal contributions for $\cB/f_a=\cW/f_a=\unit[1]{TeV^{-1}}$ with a cut of $M_{V_1 V_2} < \unit[2]{TeV}$ and $\unit[4]{TeV}$, respectively. As discussed in Sec.~\ref{sec:numerics}, the total systematic uncertainty on the signal normalization is $20\%$ (green band). The background systematics errors are taken bin-by-bin from the CMS publications (orange band).

\begin{table}[b]
\renewcommand{\arraystretch}{1.2}
\begin{tabular}{l|>{$}l<{$}@{ = }>{$}l<{$}}
\toprule
Process& \multicolumn{2}{l}{ALP EW VBS Cross Section [fb]}\\\midrule
    \multirow{2}{*}{$pp\to jjZZ$}&
    \sigma_{\rm interf.} &
    \left( 0.04 \, \cB^2 - 0.55 \, \cB \, \cW - 1.80 \, \cW^2 \right) \cdot 10^{-2} 
    \\
    &
    \sigma_{\rm signal} & 
    \left( 0.05 \, \cB^4 + 0.15 \, \cB^3 \, \cW + 1.55 \, \cB^2 \, \cW^2 + 1.66 \, \cB \, \cW^3 + 3.39 \, \cW^4 \right) \cdot 10^{-2} 
    \\\midrule
    \multirow{2}{*}{$pp\to jjZ\g$}&
    \sigma_{\rm interf.} &
    \left( 0.01 \, \cB^2 + 6.60 \, \cB \, \cW - 6.56  \, \cW^2 \right) \cdot 10^{-2} 
    \\
    &
     \sigma_{\rm signal} &
    \left( 0.19 \, \cB^4 - 0.29 \, \cB^3 \, \cW + 2.04 \, \cB^2 \, \cW^2 -2.07 \, \cB \, \cW^3 + 1.23 \, \cW^4 \right) \cdot 10^{-1} 
    \\\midrule
   \multirow{2}{*}{$pp\to jjW^\pm \g$}& 
   \sigma_{\rm interf.} &
    \cW \, \left( -1.38 \, \cB + 0.29 \, \cW \right) \cdot 10^{-3} 
    \\
   & 
   \sigma_{\rm signal} &
    \cW^2 \, \left(  5.20 \, \cB^2 + 2.12 \, \cB \, \cW + 2.81 \, \cW^2 \right) \cdot 10^{-2}  
    \\\midrule
    \multirow{2}{*}{$pp\to jjW^\pm Z$}&
    \sigma_{\rm interf.}&
    \cW \, \left( 1.15 \, \cB - 0.55 \, \cW \right) \cdot 10^{-3} 
    \\
   &
   \sigma_{\rm signal}&
    \cW^2 \,  \left(  0.90 \, \cB^2 -1.00 \, \cB \, \cW + 3.17 \, \cW^2 \right) \cdot 10^{-2}  
   \\\midrule
   \multirow{2}{*}{$pp\to jjW^\pm W^\pm$}& 
   \sigma_{\rm interf.} &
    - 0.0405 \, \cW^2 
    \\
    &\sigma_{\rm signal} &
    0.135 \, \cW^4 
    \\\bottomrule
\end{tabular}
\caption{Expected ALP EW VBS interference and pure signal cross sections at $\sqrt{s} = \unit[13]{TeV}$ as a function of the Wilson coefficients $\cW$ and $\cB$ for $f_a = \unit[1]{TeV}$ after selection cuts and requiring $M_{V_1 V_2} < \unit[4]{TeV}$. These expressions can be used to estimate the overall normalizations of the ALP signal for all distributions used in the final fits to the data.}
\label{tab:poly}
\end{table}

\clearpage
\bibliographystyle{JHEP}
\bibliography{bibliography}

\end{document}